\documentclass[12pt]{article}

\usepackage{setspace,amsthm,amsmath,amsfonts,amssymb,array}
\usepackage{graphicx,subfigure,hyperref,verbatim,authblk}
\usepackage{cite,epstopdf}
\usepackage[bottom]{footmisc}

\numberwithin{equation}{subsection}

\begin{document}
\setcounter{equation}{0}

\begin{flushright}
	NSF-KITP-14-160
\end{flushright}

\title{Minimal $3+2$ sterile neutrino model at LBNE}
\author[a]{D. Hollander\thanks{daveh@phys.psu.edu}}
\author[a,b]{I. Mocioiu\thanks{irina@phys.psu.edu}}
\affil[a]{Department of Physics, The Pennsylvania State University, University Park, PA 16802, USA}
\affil[b]{Kavli Institute for Theoretical Physics, University of California, Santa Barbara, California 93106, USA}

{\let\newpage\relax\maketitle}

\onehalfspacing

\begin{abstract}
In this paper we examine the sensitivity of the Long Baseline Neutrino Oscillation Experiment to the inclusion of  two new sterile neutrino flavors with masses in the eV range. We implement a model with a modified Casas-Ibarra parametrization which can accommodate medium scale mass eigenstates and introduces a new complex mixing angle. We explore the new mixing angle parameter space and demonstrate how LBNE can be used to either provide evidence for or rule out a particular model of sterile neutrinos. Certain three-flavor CP-violation scenarios cannot be distinguished from the sterile neutrinos. Constraints from the Daya Bay reactor experiment and T2K are used to help lift this degeneracy. 

\end{abstract}

\pagebreak

\begingroup
	\let\clearpage\relax
	\section{Introduction}
\label{sec:intro}

Precision measurements of neutrino appearance and disappearance have provided best values and tight constraints for the mixing parameters of the three flavor PMNS matrix, with the exception of the CP violating phase. Determining the value of this  phase is one of the principal goals 
of the proposed Long Baseline Neutrino Experiment (LBNE) \cite{LBNE_design} and future sensitivities have been examined in \cite{sensitivity1,sensitivity2}.  

Other experiments such as NO$\nu$A and T2K \cite{NOVA,T2K} claim CP sensitivity and can potentially aid LBNE in determining this value.  Current fits are summarized in \cite{global_fits}.  It is likely that LBNE will be constructed initially for 120 kt$\cdot$MW$\cdot$yr coming from a ~10 kt underground detector with a 1.2 MW proton beam, to be eventually increased to 600 kt$\cdot$MW$\cdot$yr from a ~40 kt detector \cite{P5}. This design is promising not only for the measurement of $\delta_{CP}$ but also for the discovery potential of new physics.

Although many oscillation phenomena can be well described by the mixing of the three active flavors of the Standard Model there have been various anomalies which cannot be accounted for, such as: $\bar{\nu}_e$ appearance excess in a $\bar\nu_\mu$ beam in LSND \cite{LSND}, an excess of electron-like events and $\bar{\nu}_e$ appearance excess in MiniBooNE  \cite{MB} which is not consistent with LSND, Gallium \cite{Gallium}, and reactor anomalies \cite{Reactor}.  The LSND anomaly, which is currently the clearest example, corresponds to a 1$\%$ $\bar{\nu}_e$ appearance probability, which cannot be accommodated by the mixing of the three standard neutrino flavors.

One possible explanation for this anomaly is the existence of new, sterile (right handed) neutrinos with a new mass scale $\Delta m^2 \sim 1$ eV$^2$.  In general, $3+n$ models of sterile neutrinos can be phenomenological where the mass matrix is generic, or minimal extensions of the Standard Model where only $n$ Weyl singlets are added.  Different models of sterile neutrinos have been examined, including $3+1$ (1 new neutrino with $\Delta m^2 \sim 1$ eV$^2$) minimal\cite{3+1Rev1} and phenomenological \cite{3+2Rev1} models , $3+2$ (2 new neutrinos with $\Delta m^2 \sim 1$ eV$^2$) phenomenological models\cite{3+2Rev1,3+2Rev2}, and other models such as $3+1+1$ (1 new neutrino with $\Delta m^2 \sim 1$ eV$^2$, 1 additional new neutrino with $\Delta m^2 \gg 1$ eV$^2$) minimal \cite{Sterile1} and phenomenological \cite{Sterile2,Sterile3} models.  For $3+n$ minimal models a simpler mixing parameterization of this model can be found from Casas-Ibarra \cite{Casas_Ibarra} which assumes a decoupling between heavy and light neutrinos; this parameterization breaks down when $\Delta m_s^2 \sim 1$ eV$^2$.  Generalizations of the Casas-Ibarra parameterization, valid in all parameter space, have been proposeded in \cite{Blennow} and \cite{Mod_Casas_Ibarra}. 

Light sterile neutrinos affect the oscillations of active flavors and can impact results of standard oscillation experiments as well as particle-astrophysics and cosmology.  The impact on cosmology from phenomenological \cite{Lepton_asymm1,Lepton_asymm2,Lepton_asymm3,Lepton_asymm4} and minimal \cite{Lepton_asymm5,Lepton_asymm6} models have been studied.  The impact of a minimal model on astrophysics has been examined in \cite{Astro}.

In this paper we study a $3+2$ minimal extension of the Standard Model including sterile neutrinos where the mixing is parameterized with a modified Casas-Ibarra method implemented in \cite{Mod_Casas_Ibarra}, which introduces one new mass scale and one complex mixing angle.  The relatively small number of new parameters of this minimal model allows us to better constrain them from existing experiments as well as make predictions for LBNE.  Using this parameterization we can compute the appearance probabilities relevant for LBNE with different values of the 3+2 mixing parameters and compute future sensitivities for sterile neutrino discovery at LBNE.  We find that for certain mixing parameter values there are degeneracies between the 3-flavor and 3+2 flavor descriptions.  We use reactor data from Daya Bay, which is not sensitive to $\delta_{CP}$, and data from T2K to provide further constraints to help lift the degeneracies. These results, when confronted with real data from LBNE, can help confirm or rule out this particular model of sterile neutrinos.

	\graphicspath{{D:/Physics/Neutrinos/Plots/LBNE/}}

\section{3+2 model and mixing parameterization}
\label{sec:model}

Extending the Standard Model is performed by adding Weyl fields that are invariant under Standard Model gauge transformations.  
An arbitrary sterile neutrino model can be constructed by adding in Weyl fields in such a way that they pair up to form more than three Dirac neutrinos.  In this model, the minimal extension of the Standard Model consists of only adding in two $SU(2)_L$ singlet Weyl fields.  The renormalizable Lagrangian for sterile and active neutrino mixing is given by
\begin{equation}
	\mathcal{L} \supset -\bar{l}^\alpha_L Y^{\alpha j}\Phi \nu^j_R - \frac{1}{2} \bar{\nu}_R^{ic}M_R^{ij}\nu_R^j + h.c.
\end{equation}
and a basis is chosen such that the neutrino mass matrix is given by
\[
	M_\nu = \begin{pmatrix}
		0 & m_Y \\
		m_Y^T & M_R 
	\end{pmatrix}
\]
where $m_Y$ is the $3 \times 2$ Yukawa mass matrix, and $M_R=\textrm{Diag}(m_4,m_5)$ where $m_4$ and $m_5$ are the masses of the two new mass eigenstates; in this model $m_4 \approx m_5 \sim 1$ eV.

A modified version of the Casas-Ibarra mixing parameterization for this model was studied in \cite{Mod_Casas_Ibarra}.  The full Casas-Ibarra parameterization factorizes heavy and light neutrino mixing by using an expansion in $m_Y/M_R$ that is truncated after first order \cite{Casas_Ibarra}, which is insufficiently precise for this model where sterile masses are $\sim 1$ eV.  In a generic $SU(5)$ theory of active and sterile neutrino flavors, there are 10 mixing angles, and 9 phases.  A nice feature of the minimal model is that it greatly reduces the number of mixing angles and phases 
describing the model, the details can be seen in table \ref{tab:DOF}.  The motivation for choosing this particular model is that it is the 
minimal 
extension of the Standard Model including sterile neutrinos which can still explain existing oscillation data \cite{Mod_Casas_Ibarra}.
\begin{table}
	\centering
	\begin{tabular}{l|c|c|c} \hline
		Model & \# $\Delta m^2$ & \# Angles & \# Phases\\ \hline
		3$\nu$ & 2 & 3 & 1 \\
		3+2 MM & 4 & 4 & 3 \\
		3+2 PM & 4 & 9 & 5 \\ \hline
	\end{tabular}
	\caption{Number of mass, angle, and phase parameters.  PM refers to the phenomenological model which contains 
	more than two additional right handed Weyl fields, while MM refers to the minimal model used in this paper.
	\label{tab:DOF}}
\end{table}

Setting $m_1 = 0$ the mass eigenstates become $m_i=\sqrt{\Delta m_{i1}^2}$; eliminating the zero mass eigenstate in the matrix reduces the dimension of the mixing matrix that needs to be considered.  
The full unitary mixing matrix for the five flavor system can be written as: 
\[
	U=\begin{pmatrix}
		U_{aa} & U_{as} \\
		U_{sa} & U_{ss} \end{pmatrix}
\]
where 
\[
	\begin{aligned}
		U_{aa} &= U_{PMNS}\begin{pmatrix}
			1 & 0 \\
			0 & H \end{pmatrix} \\
		U_{as} &= iU_{PMNS} \begin{pmatrix}
			0 \\
			Hm_l^{1/2}R^\dagger M_h^{-1/2} \end{pmatrix} \\
		U_{sa} &= \begin{pmatrix} 0 & \bar{H}M_h^{-1/2}Rm_l^{1/2} \end{pmatrix} \\
		U_{ss} &= \bar{H}.
	\end{aligned}
\]
The matrices are given by
\[
	\begin{aligned}
		R &= \begin{pmatrix}
			\cos(\theta_{45}+i\gamma_{45}) & \sin(\theta_{45}+i\gamma_{45}) \\
			-\sin(\theta_{45}+i\gamma_{45}) & \cos(\theta_{45}+i\gamma_{45}) \end{pmatrix} \\
		m_l &= \textrm{Diag}(m_2, m_3) \\
		M_h &= \textrm{Diag}(m_4, m_5) \\
		H^{-2} &= I+m_l^{1/2}R^\dagger M_h^{-1}Rm_l^{1/2} \\
		\bar{H}^{-2} &= I+M_h^{-1/2}Rm_l R^\dagger M_h^{-1/2}.
	\end{aligned}
\]
Here we have introduced the complex mixing angle $z_{45}=\theta_{45}+i\gamma_{45}$ which explicitly mixes the new, sterile mass eigenstates.  It should be emphasized that $\gamma_{45}$ is not a phase itself, but rather a purely imaginary component of $z_{45}$.  $U_{PMNS}$ looks like an usual $3\nu$ mixing matrix given by
\[
	U_{PMNS} = \begin{pmatrix}
		1 & 0 & 0 \\
		0 & c_{23} & s_{23} \\
		0 & -s_{23} & c_{23} \end{pmatrix}
	\begin{pmatrix}
		c_{13} & 0 & s_{13}e^{-i\delta} \\
		0 & 1 & 0 \\
		-s_{13}e^{i\delta} & 0 & c_{13} \end{pmatrix}
	\begin{pmatrix}
		c_{12} & s_{12} & 0 \\
		-s_{12} & c_{12} & 0 \\
		0 & 0 & 1 \end{pmatrix}
	\begin{pmatrix}
		1 & 0 & 0 \\
		0 & 1 & 0 \\
		0 & 0 & e^{i\alpha} \end{pmatrix}.
\]
It was found in \cite{Mod_Casas_Ibarra} that this 
model provided best fits to existing oscillation data for the normal hierarchy, therefore we exclusively work in the normal hierarchy in this paper. The mixing angles $\theta_{13},\theta_{23}$,and $\theta_{12}$, as well as the mass splittings $m_2^2\equiv \Delta m_{21}^2$ and $m_3^2\equiv\Delta m_{31}^2$ are fixed to global fits from \cite{global_fits}.  The sterile mass eigenvalues are fixed to their best fit values in the phenomenological 3+2 model for the normal hierarchy: $m_4^2\equiv\Delta m_{41}^2=0.47$ eV$^2$ and $m_5^2\equiv\Delta m_{51}^2=0.87$ eV$^2$.

LBNE is sensitive to $\delta_{CP}$ and $\gamma_{45}$, the imaginary component of the sterile mass mixing angle, therefore both of these parameters are allowed to vary for the LBNE probabilities in sections \ref{sec:model} and \ref{sec:LBNE_sensitivity}, all other mixing parameters are fixed to best fit values.  This will allow us to assess the ability of LBNE to distinguish a signal from sterile neutrinos over the signal from the three flavor scenario.

Daya Bay is sensitive to $\gamma_{45}$ in addition to $\theta_{13}$; in section \ref{sec:DB_sensitivity} these parameters will be allowed to vary to help constrain the allowed values of $\gamma_{45}$. 
As before, all other mixing parameters will be fixed to the best fit values of \cite{global_fits,Mod_Casas_Ibarra}.

LBNE measures $\nu_e$ appearance from the initially produced $\nu_{\mu}$ at a baseline of $L=1300$ km.  Examples of oscillation probabilities $P(\nu_{\mu} \to \nu_e)$ and $P(\bar{\nu}_{\mu} \to \bar{\nu}_e)$ at LBNE when only $\delta_{CP}$ is varied and all other parameters are fixed to their best fit values can be seen in Figs. \ref{fig:CP} and \ref{fig:CP_anti}.
\begin{figure}[ht!!!]
	\centering
	\subfigure[]{\includegraphics[width=0.45\textwidth]{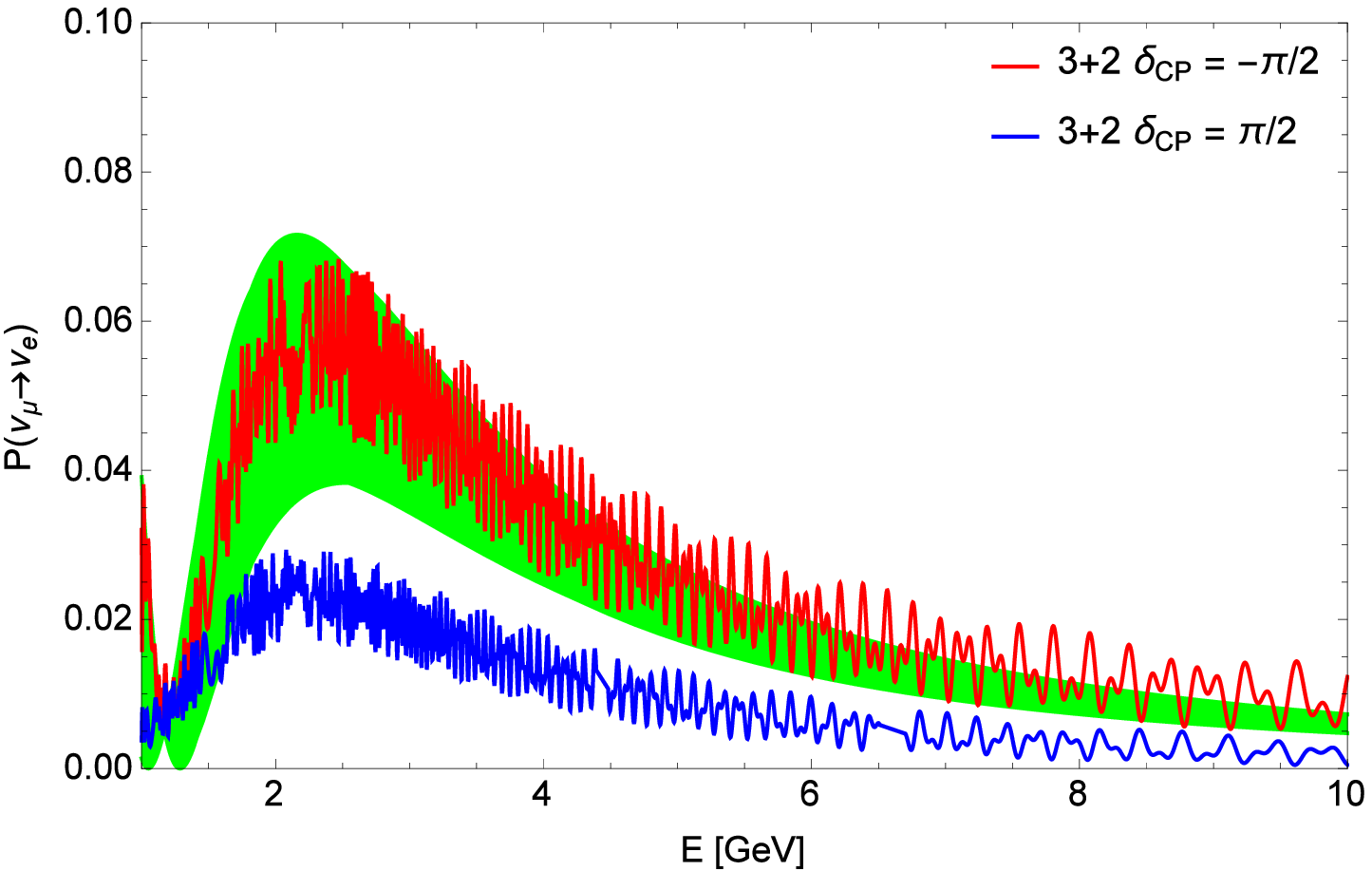}\label{fig:CP}}
	\subfigure[]{\includegraphics[width=0.45\textwidth]{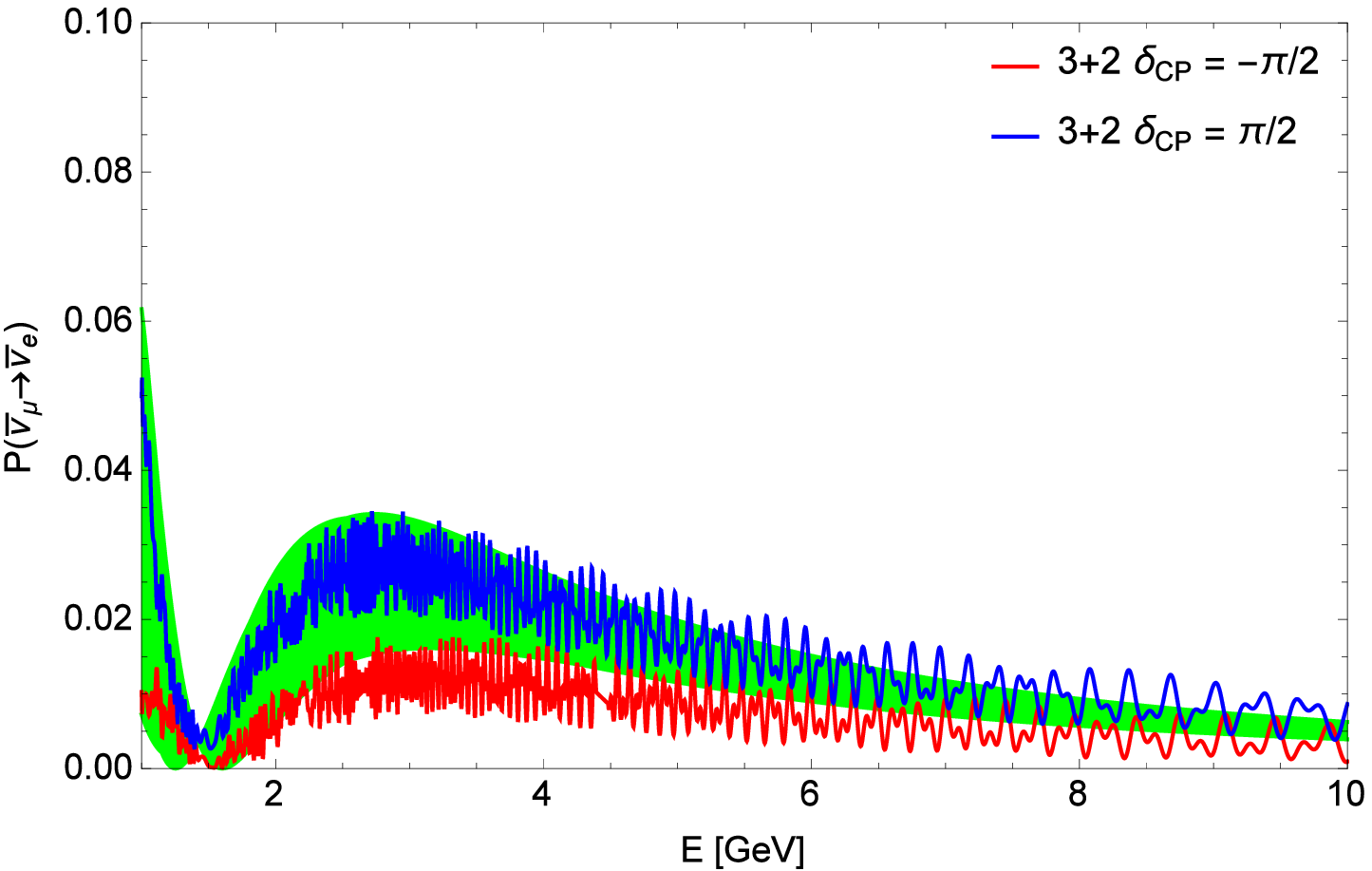}\label{fig:CP_anti}}
	\caption{(Color online) Plots of the probabilities $P(\nu_{\mu} \to \nu_e)$ (left) and $P(\bar{\nu}_{\mu} \to 
	\bar{\nu}_e)$ (right) where the solid green bands are the three-flavor probabilities over the range of 
	$\delta_{CP} \in \left[ -\pi, \pi \right]$; red and blue curves are probabilities in the 
	minimal 3+2 model.  All other unspecified mixing parameters are fixed to the best fit values from 
	\cite{Mod_Casas_Ibarra}.  The rapid, small oscillations present in the 3+2 probabilities are controlled by 
	$\theta_{45}$, which mixes the two sterile mass eigenstates.}
\end{figure}
In both plots the solid green bands are the three-flavor probabilities over the range of $\delta_{CP} \in \left[ -\pi, \pi \right]$; red and blue curves are probabilities in the minimal 3+2 model.  Unlike in the 3+2 scheme, the three-flavor probabilities are smooth curves.  The 3+2 probabilities exhibit rapid oscillations due to the large mass scale of the sterile states: $m_4 \sim m_5 = \mathcal{O}(\textrm{eV})$.  The amplitude of these oscillations is controlled by $\theta_{45}$, which mixes the two sterile mass states.  Given the energy resolution of LBNE and Daya Bay, such fast oscillations cannot be resolved.  We therefore fix $\theta_{45}$ to the normal hierarchy best fit value in \cite{Mod_Casas_Ibarra}.  It can be seen that for certain values of $\delta_{CP}$ the probabilities in the 3+2 scheme lie outside of the allowed region for three-flavors, which can lead to distinguishable signals.  For experiments with a sufficiently long baseline, such as LBNE and Daya Bay, where $\Delta m_{21}^2$ mixing is negligible, the upper left $3\times 3$ sub-matrix of the larger, $5\times 5$ mixing matrix is just the PMNS matrix to leading order.  Therefore profile differences between the three-flavor and 3+2 flavor scenarios due to $\delta_{CP}$ must be a sub-leading order effect.

Examples of $P(\nu_{\mu} \to \nu_e)$ and $P(\bar{\nu}_{\mu} \to \bar{\nu}_e)$ at LBNE when $\delta_{CP}$ and $\gamma_{45}$ are varied independently and all other parameters are fixed to their best fit values can be seen in Figs. \ref{fig:CP_G45_neg}, \ref{fig:CP_G45_pos}, \ref{fig:anti_CP_G45_neg}, and \ref{fig:anti_CP_G45_pos}.  
\begin{figure}[ht!!!]
	\centering
	\subfigure[]{\includegraphics[width=0.45\textwidth]{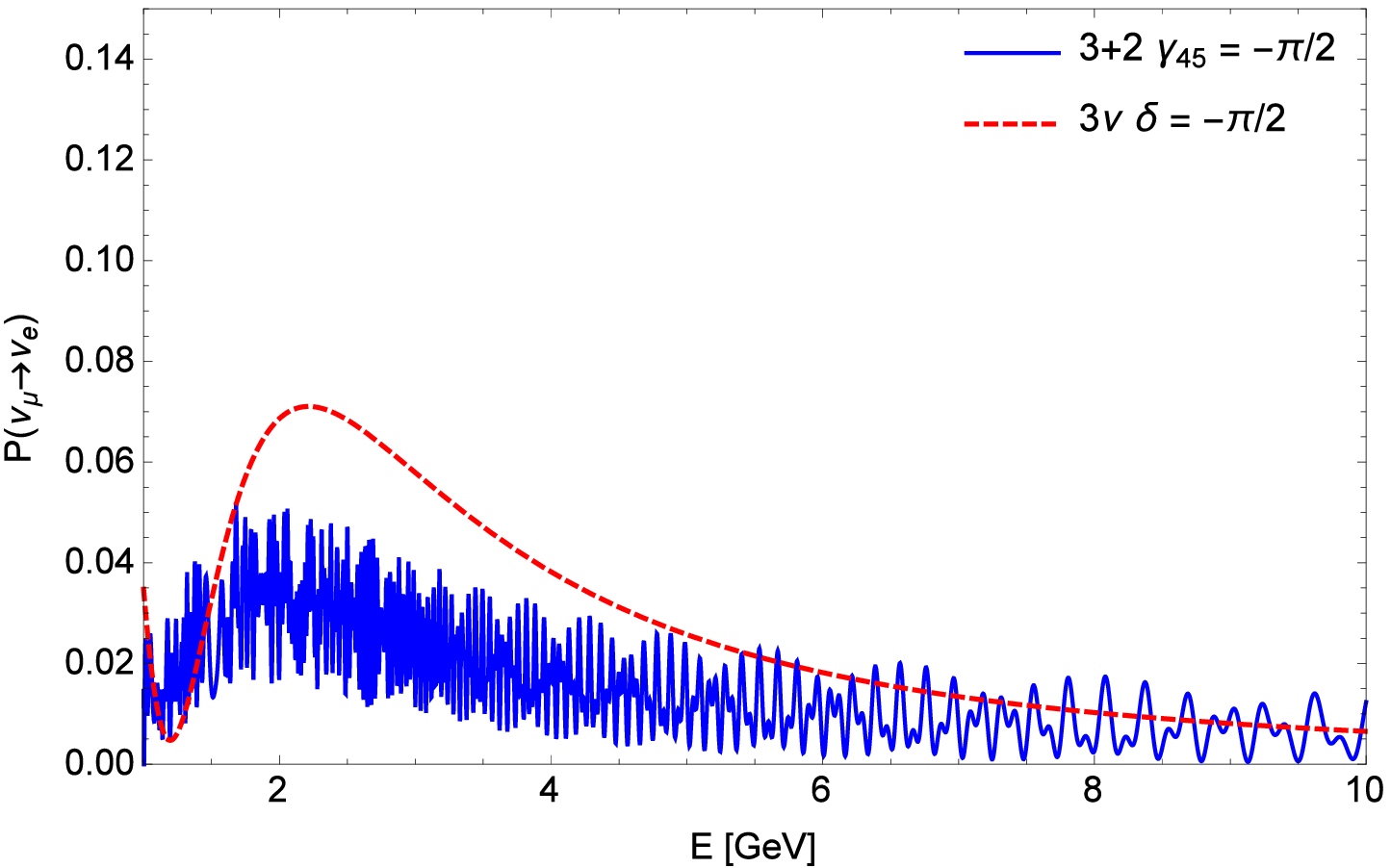}\label{fig:CP_G45_neg}}
	\subfigure[]{\includegraphics[width=0.45\textwidth]{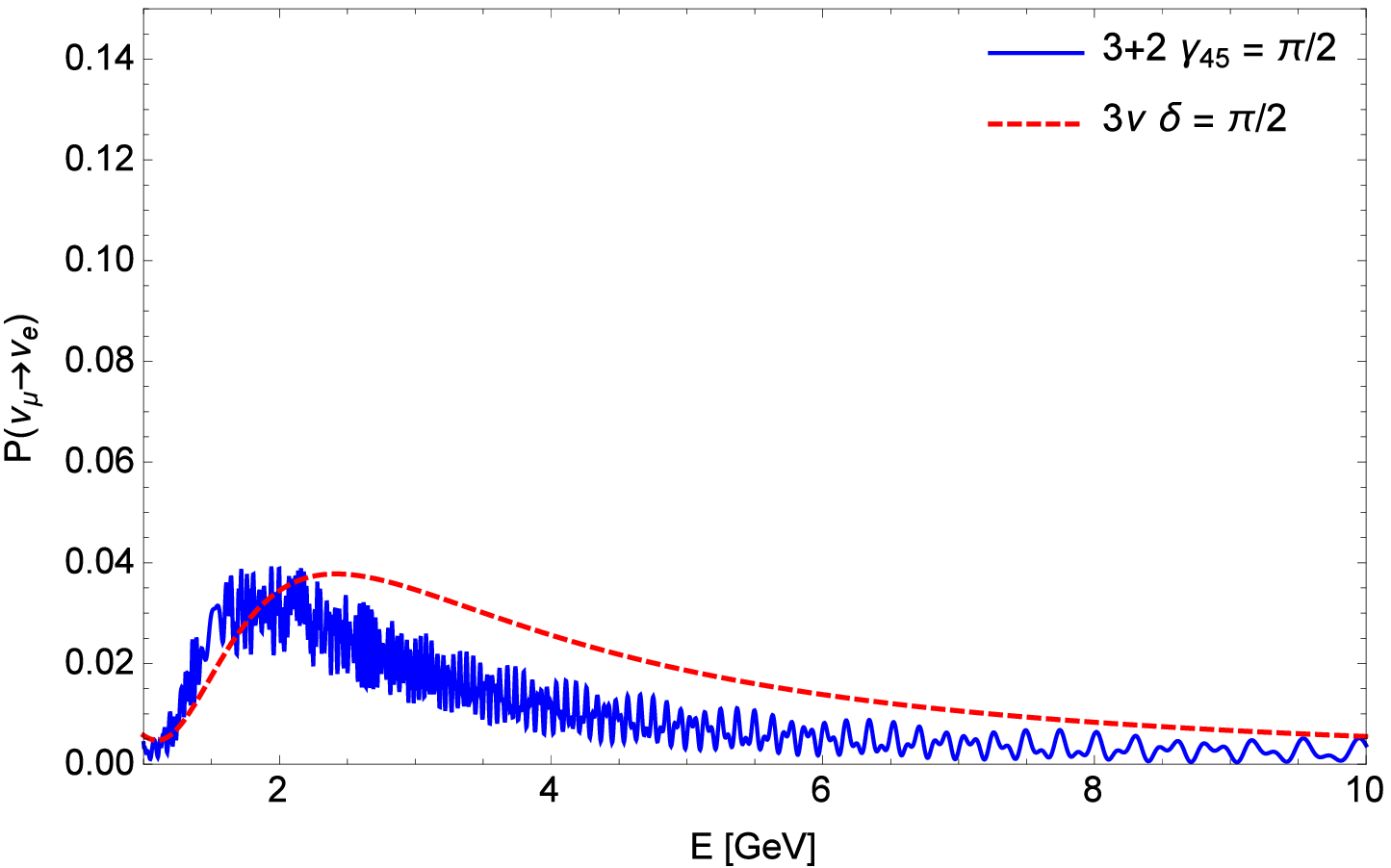}\label{fig:CP_G45_pos}}\\
	\subfigure[]{\includegraphics[width=0.45\textwidth]{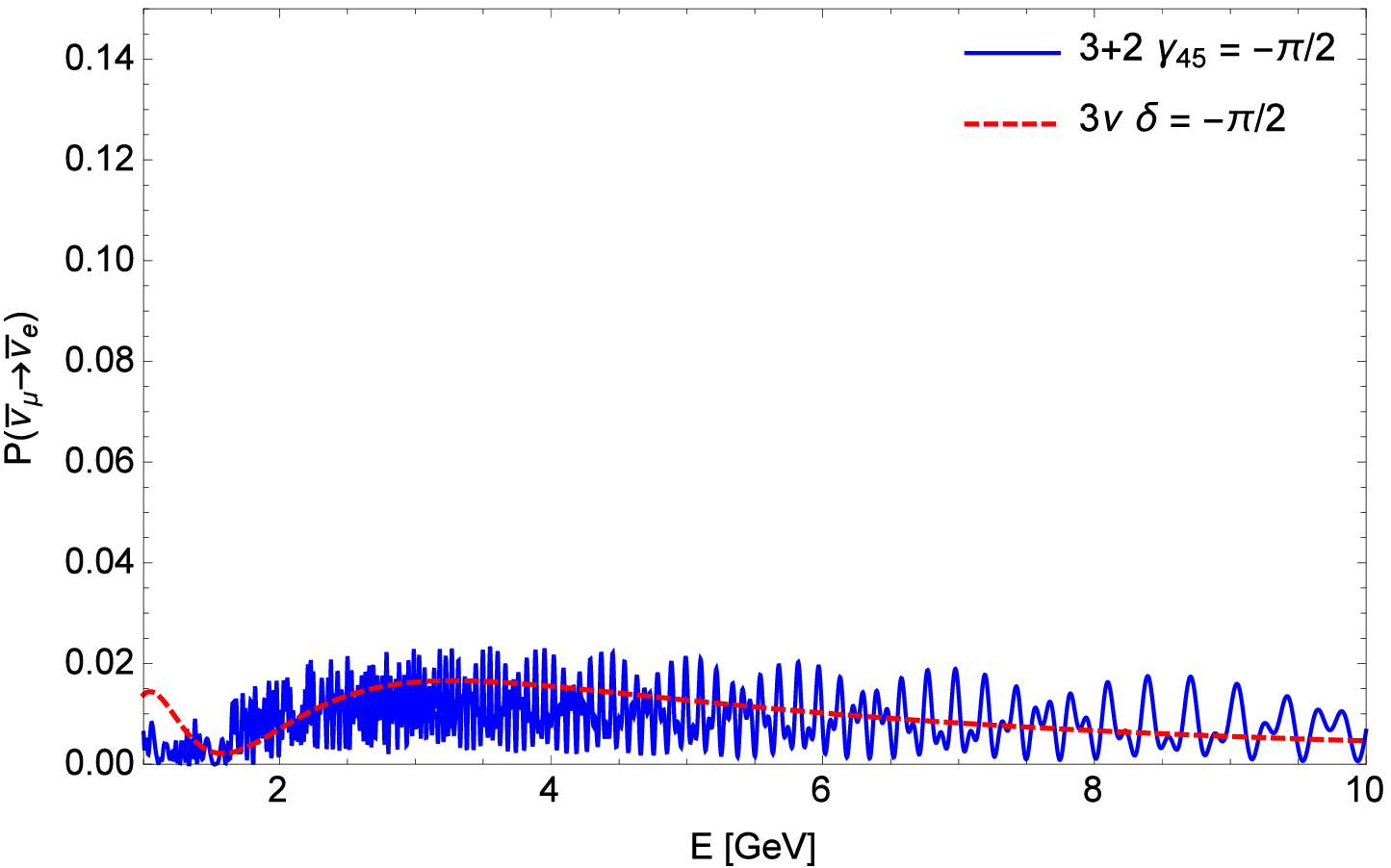}\label{fig:anti_CP_G45_neg}}
	\subfigure[]{\includegraphics[width=0.45\textwidth]{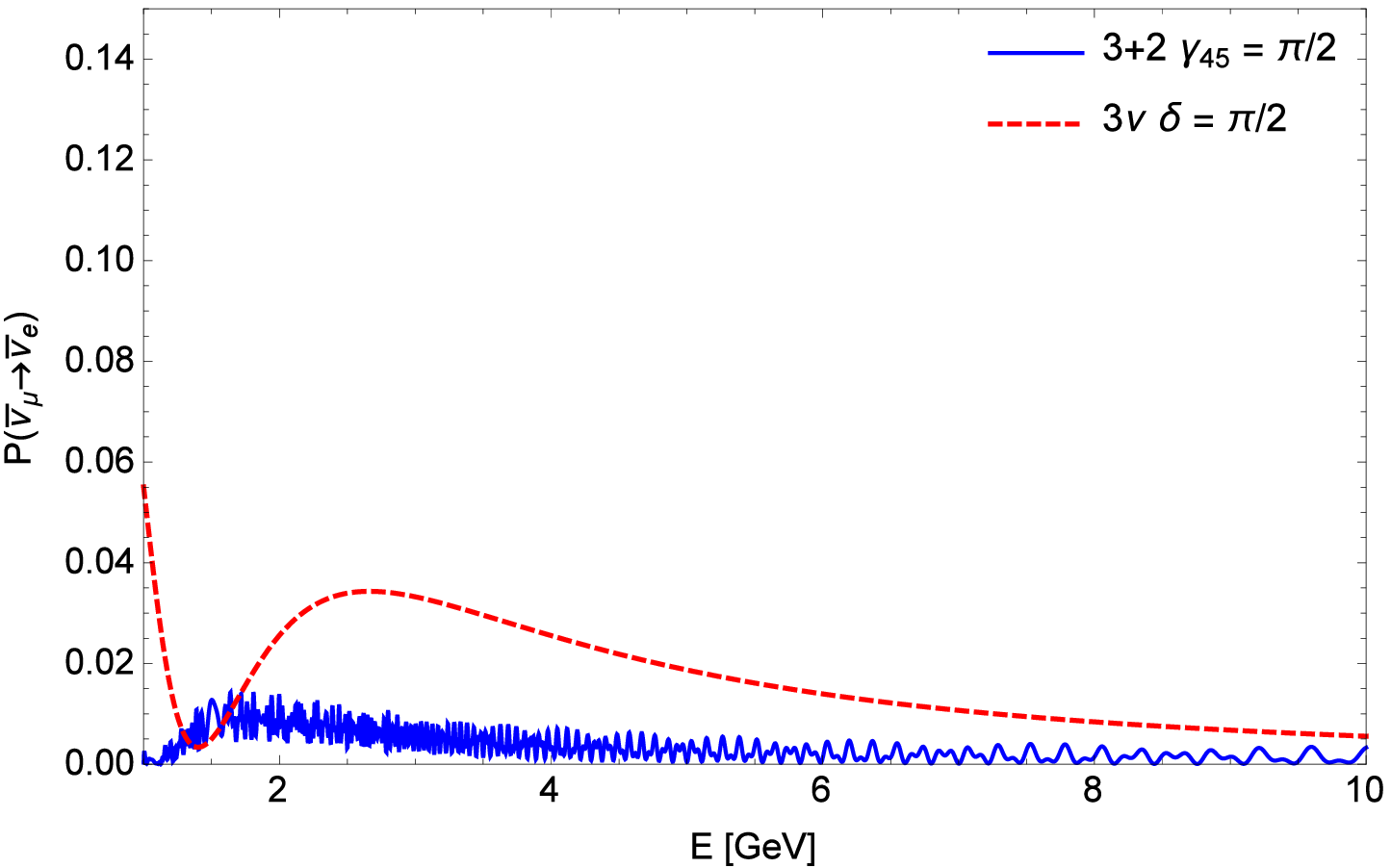}\label{fig:anti_CP_G45_pos}}
	\caption{(Color online) (Top) Plots of the probabilities $P(\nu_{\mu} \to \nu_e)$, and (bottom) Plots of 
	the probabilities $P(\bar{\nu}_{\mu} \to \bar{\nu}_e)$ for (left) $\delta_{CP} = \gamma_{45}=-\pi/2$ and (right) 
	$\delta_{CP} = \gamma_{45}=\pi/2$.  All other, unspecified mixing parameters are fixed to the best fit values from 
	\cite{Mod_Casas_Ibarra}.  Dotted curves are for three flavors, while solid curves are probabilities in the 3+2 model.}
\end{figure}
In each plot we take $\gamma_{45}$ and $\delta_{CP}$ to have the same value for purpose of example only.  It can be seen how certain values of $\gamma_{45},\delta_{CP}$ lead to probabilities that are difficult to distinguish, while other values result in clearly separated profiles.

The results from Figs. \ref{fig:CP}, \ref{fig:CP_anti} and \ref{fig:CP_G45_neg} - \ref{fig:anti_CP_G45_pos} may indicate that for LBNE it may be difficult to distinguish the signals coming from Standard Model mixing and minimal extra mixing with sterile neutrinos for certain values of $\delta_{CP}$.  However for other CP-violating phases LBNE is poised to provide evidence for sterile neutrinos in this 3+2 minimal extension, or help to rule out this model for active-sterile mixing.
	\graphicspath{{D:/Physics/Neutrinos/Plots/LBNE/}}

\section{Sterile sensitivity}
\label{sec:Sensitivity}

\subsection{Sensitivity in LBNE}
\label{sec:LBNE_sensitivity}

LBNE has a 1300 km baseline, so matter effects become important for this experiment.  The neutrino and anti-neutrino effective Hamiltonians are given by
\begin{equation}
	\begin{aligned}
		H^{\textrm{eff}}_{\nu} &= \frac{\Delta m_{31}^2}{2E} U \textrm{ diag}(0,\alpha,1,\beta,\gamma) U^\dagger 
		+ \frac{\sqrt{2}G_F n_e}{2} \textrm{ diag}(1,-1,-1,0,0) \\
		H^{\textrm{eff}}_{\bar{\nu}} &= \frac{\Delta m_{31}^2}{2E} U^* \textrm{diag}(0,\alpha,1,\beta,\gamma) U^{t} 
		+ \frac{\sqrt{2}G_F n_e}{2} \textrm{ diag}(-1,1,1,0,0)
	\end{aligned}
\end{equation}
where $U$ is the $5\times 5$ mixing matrix, $\alpha\equiv \Delta m_{21}^2/\Delta m_{31}^2,\beta\equiv \Delta m_{41}^2/\Delta m_{31}^2,\gamma\equiv \Delta m_{51}^2/\Delta m_{31}^2$.  The matter terms result from the fact that neutral current interactions, in addition to charged current, now provide a relative phase in the flavor evolutions in the presence of sterile neutrinos.  The large terms $\beta, \gamma$ give rise to rapid oscillations which have periods much narrower than the LBNE energy resolution.  LBNE is therefore insensitive to the large mass scale oscillation terms that are essentially averaged out.  Expanding the matrix terms it can be shown that LBNE is mainly sensitive to the mixing matrix elements $U_{e2},U_{e3},U_{\mu 2},U_{\mu 3},U_{\tau 2},U_{\tau 3},U_{s_1 2},U_{s_1 3},U_{s_2 2},U_{s_2 3}$, and it can provide constraints to some of the mixing parameters included in those elements.

The number of events in each energy bin expected at LBNE is calculated as
\begin{equation}
	N_i=\int_{\Delta E_i} dE \left(\phi^{CC}_{\nu}(E)P_{\nu_{\mu}\nu_e}(E,\delta_{CP},\gamma_{45}) + 
	\phi^{CC}_{\bar{\nu}}(E)P_{\bar{\nu}_{\mu}\bar{\nu}_e}(E,\delta_{CP},\gamma_{45}) \right)
	\label{eq:events}
\end{equation}
where $\Delta E_i \equiv 15\% /\sqrt{E(\textrm{GeV})}$ is the energy resolution of LBNE \cite{LBNE_design}.  $\phi^{CC}$ is the energy-dependent charged-current rate, the rate is taken from the LBNE design report \cite{LBNE_design} and GLoBES \cite{globes1,globes2}.  An example of the number of events with sterile neutrinos in the full LBNE design, after one year of data and the comparison to the three-flavor predictions can be seen in Figs. \ref{fig:anti_events} and \ref{fig:anti_events+180deg}.
\begin{figure}[ht!!!]
	\centering
	\subfigure[]{\includegraphics[width=0.45\textwidth]{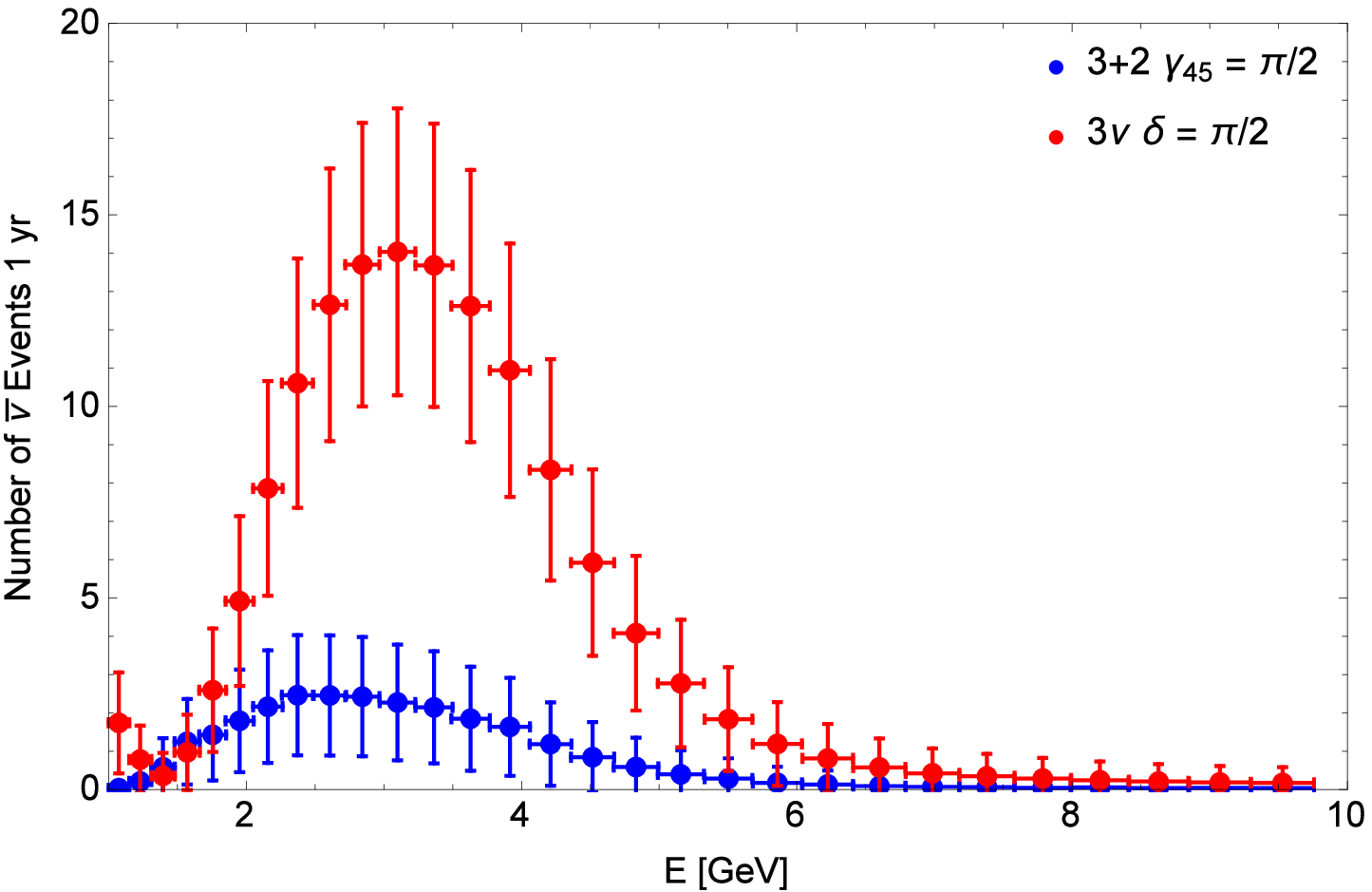}\label{fig:anti_events}}
	\subfigure[]{\includegraphics[width=0.45\textwidth]{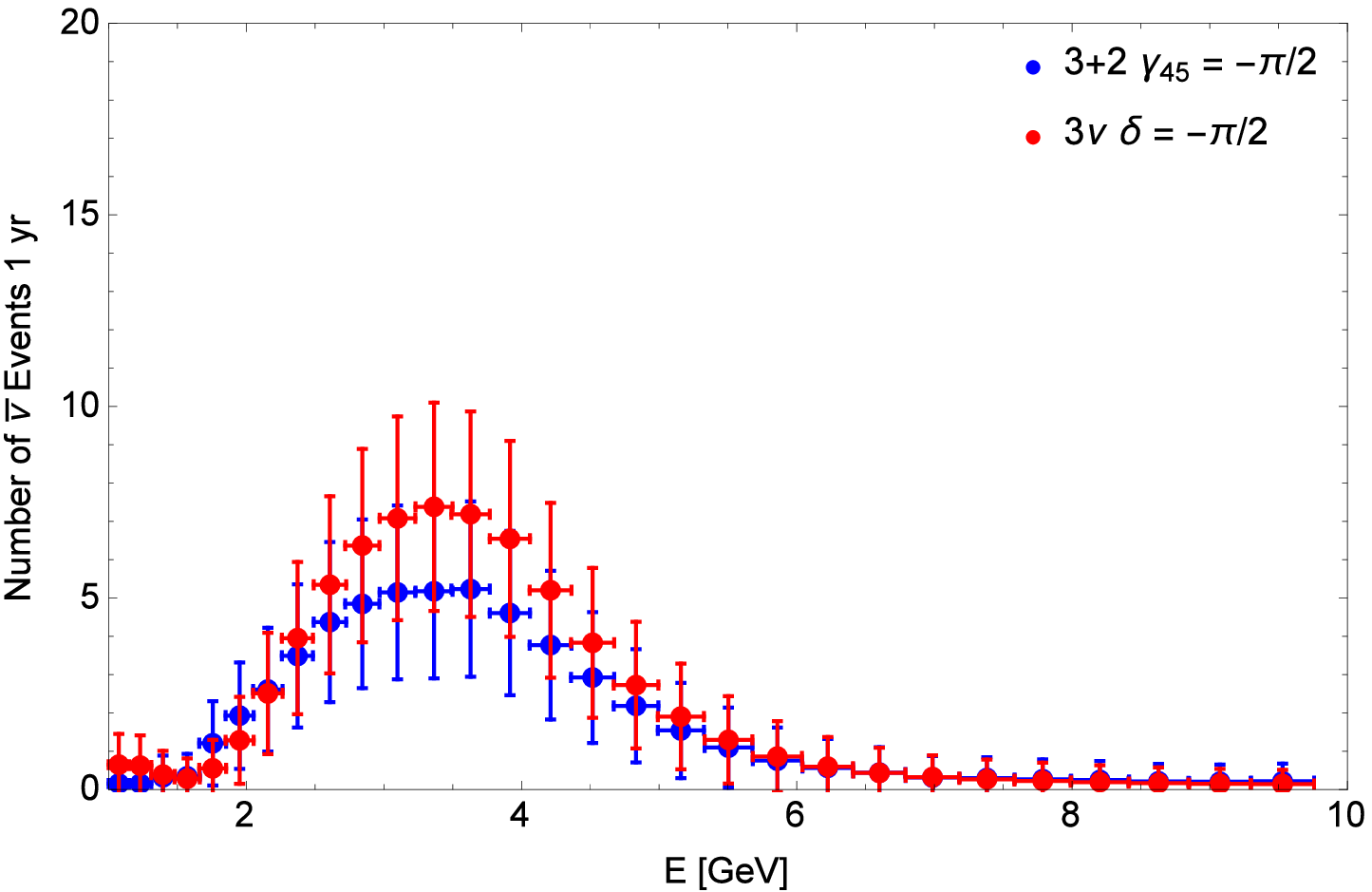}\label{fig:anti_events+180deg}}
	\caption{(Color online) The number of anti-neutrino events in 40 kt LBNE after 1 year for (left) $\delta_{CP}=
	\gamma_{45}=\pi/2$ and (right) $\delta_{CP}=\gamma_{45}=-\pi/2$.  Error bars are statistical only; this width 
	of each bin is the LBNE energy resolution $\Delta E_i \equiv 15\% /\sqrt{E(\textrm{GeV})}$.}
\end{figure}
For certain combinations of $\delta_{CP}$ and $\gamma_{45}$ there is little difference between the number of events in each energy bin expected with or without sterile neutrinos, while other parameter values can lead to a clear deficit resulting from oscillations into sterile flavors.
\begin{figure}[ht!!!]
	\centering
	\subfigure[]{\includegraphics[width=0.45\textwidth]{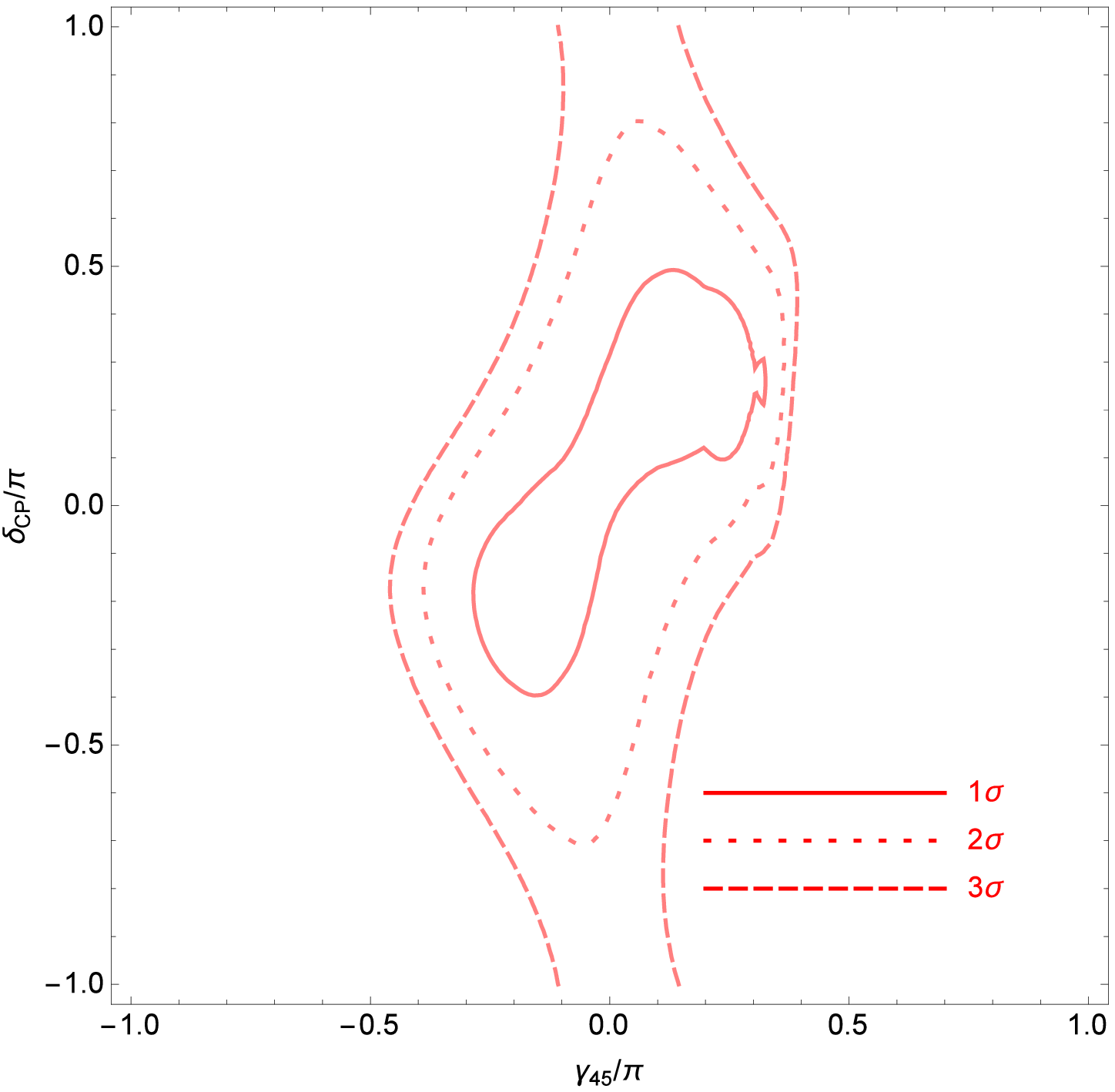}\label{fig:sensitivity_initial}}
	\subfigure[]{\includegraphics[width=0.45\textwidth]{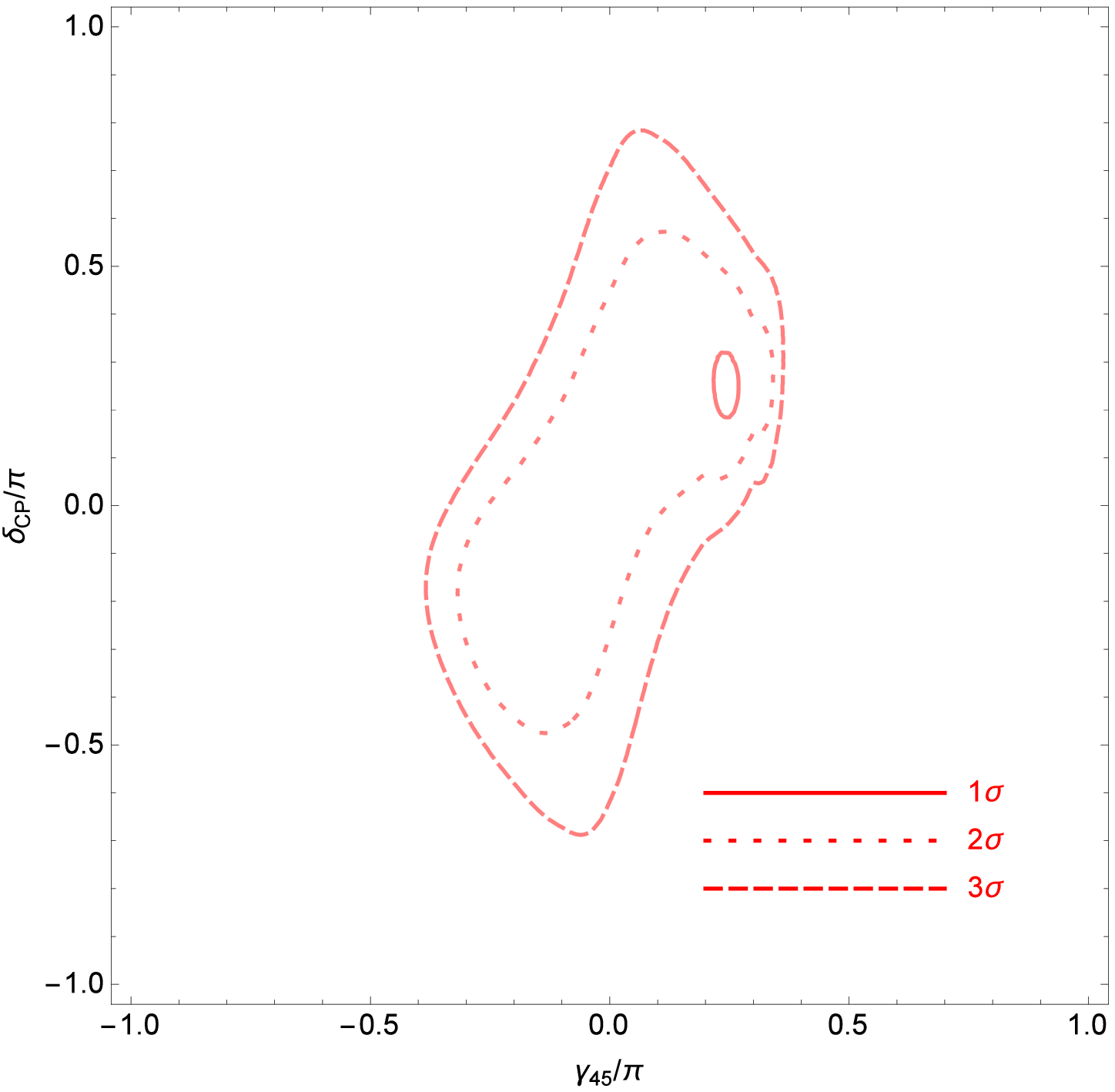}\label{fig:sensitivity_twice_initial}} \\
	\subfigure[]{\includegraphics[width=0.45\textwidth]{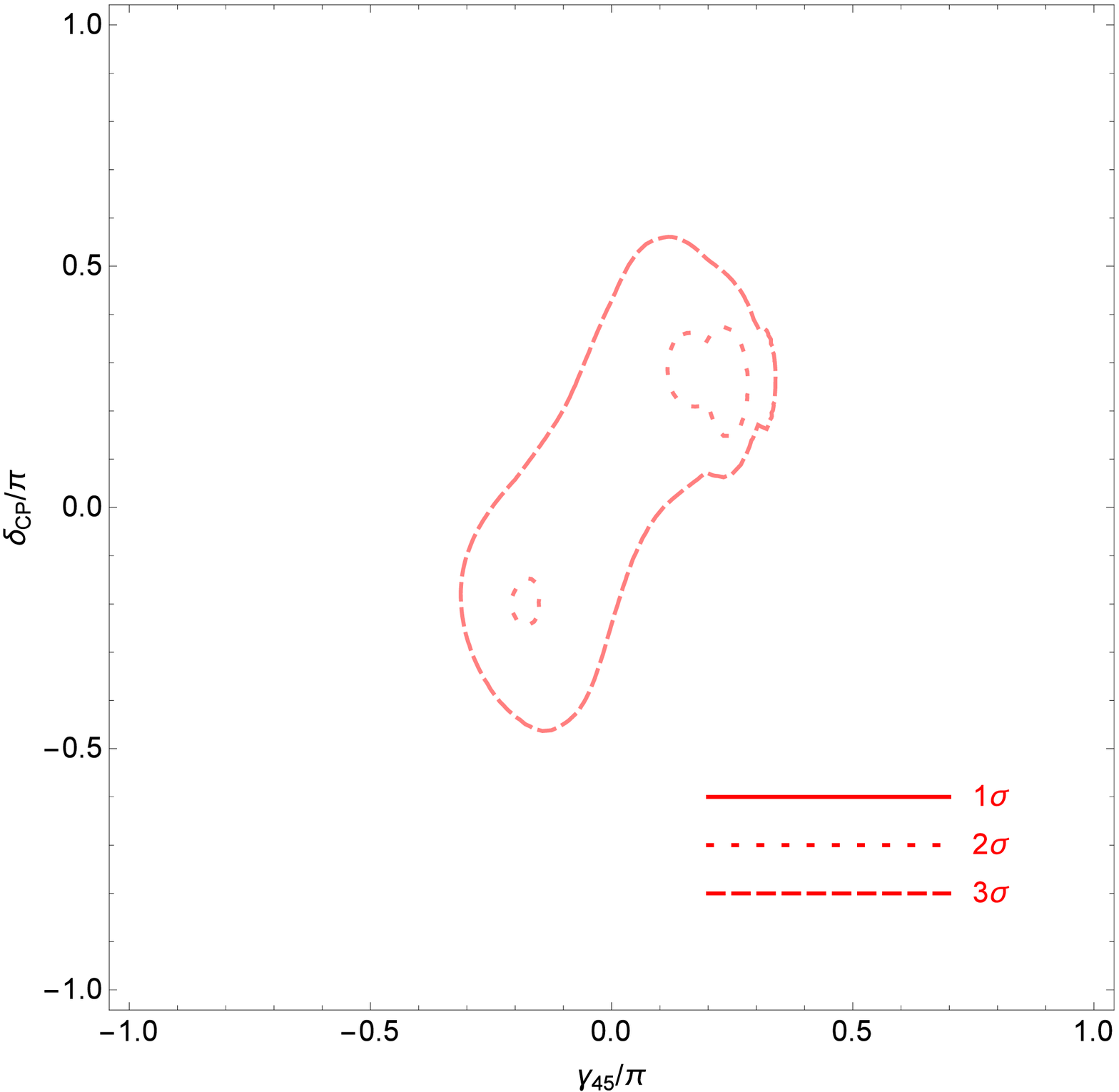}\label{fig:sensitivity_full}}
	\caption{(Color online) (Top) Sensitivity contours for sterile neutrino discovery in LBNE at 10 kt (left)
	and 20 kt (right), and the full, 40 kt design (bottom) all after $5\nu + 5\bar{\nu}$ years of data taking.  The 
	$3\sigma$ contour cannot be closed after 5 + 5 years with the 10 kt detector size, therefore distinguishing the two 
	signals cannot be done at $3\sigma$ for the initial design.}
\end{figure}

In order to quantify the ability for LBNE to distinguish between signatures from only three-flavors and 3+2 flavors we define the sensitivity quantity
\begin{equation}
	\Delta \chi^2 = \sum_{i\in I}\frac{\left(N_{\nu+\bar{\nu}}^{3+2}(E_i,\delta_{CP},\gamma_{45}) - 
	N_{\nu+\bar{\nu}}^{3\nu}(E_i,\delta_{CP})\right)^2}{N_{\nu+\bar{\nu}}^{3\nu}(E_i,\delta_{CP})}
	\label{eq:Chi2}
\end{equation}
where $I$ is a set of energy bins indices.  There is a $\sim 3.6\%$ total systematic uncertainty \cite{LBNE_design} in $\nu_e$ appearance which arises from beam flux normalization uncertainties, cross-section uncertainties as well as energy scale uncertainties.  The flux normalization uncertainty is highly correlated between $\nu_{\mu}$ and $\nu_e$.  The number of events in the 40 kt LBNE design after one year of data taking is small, on the order of $\sim 10$ per energy bin at the peak, and therefore the total systematic uncertainty has a small affect on the sensitivity contours.  Including the LBNE total systematic uncertainty for $\nu_e$ appearance would have only a small effect on the $\Delta \chi^2$ values over the entire parameter space, and decreasing the systematic uncertainties over time will not change the sensitivity greatly.
\begin{figure}[ht!!!]
	\centering
	\subfigure[]{\includegraphics[width=0.45\textwidth]{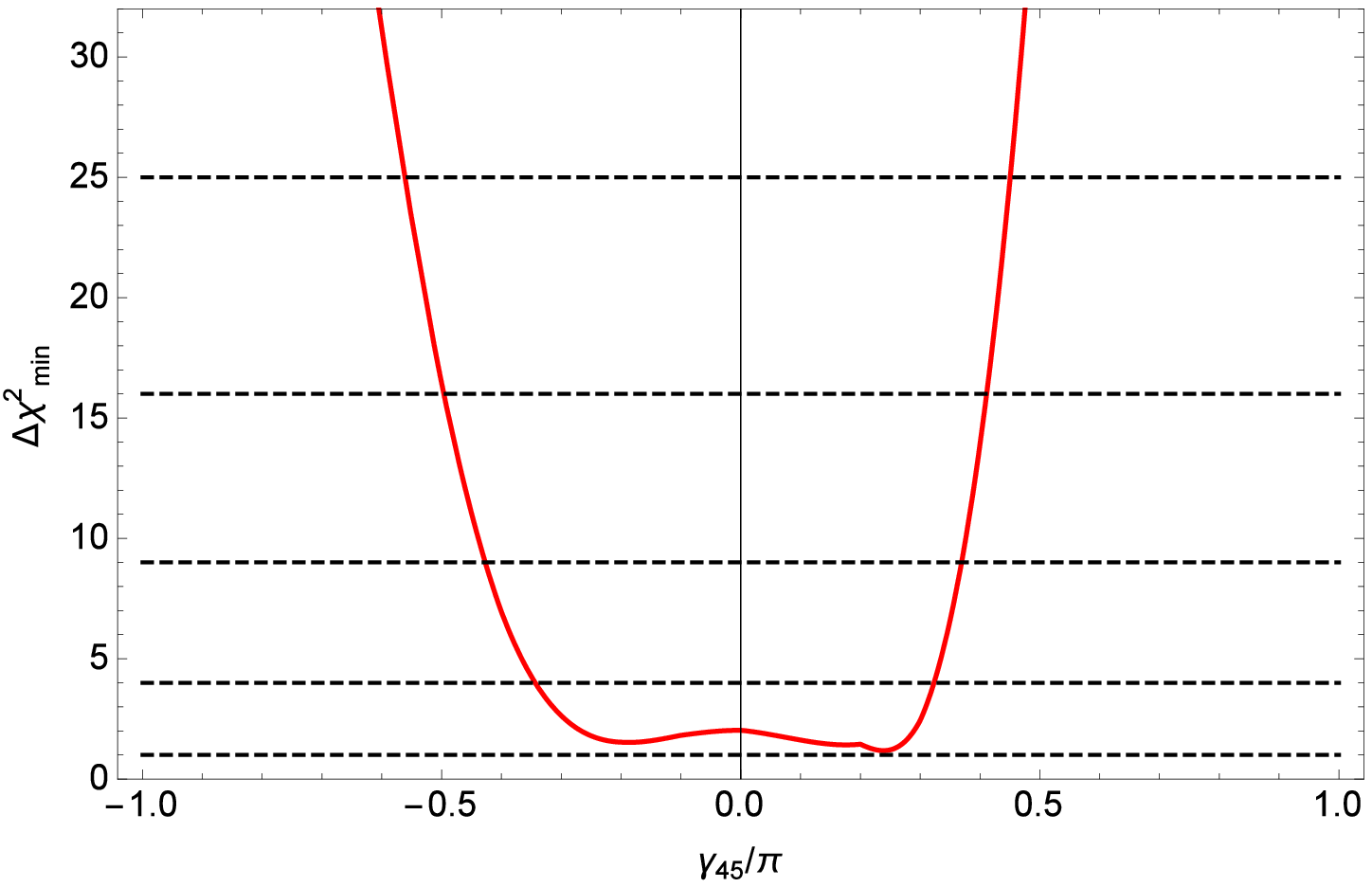}\label{fig:minChi2G45_initial}}
	\subfigure[]{\includegraphics[width=0.45\textwidth]{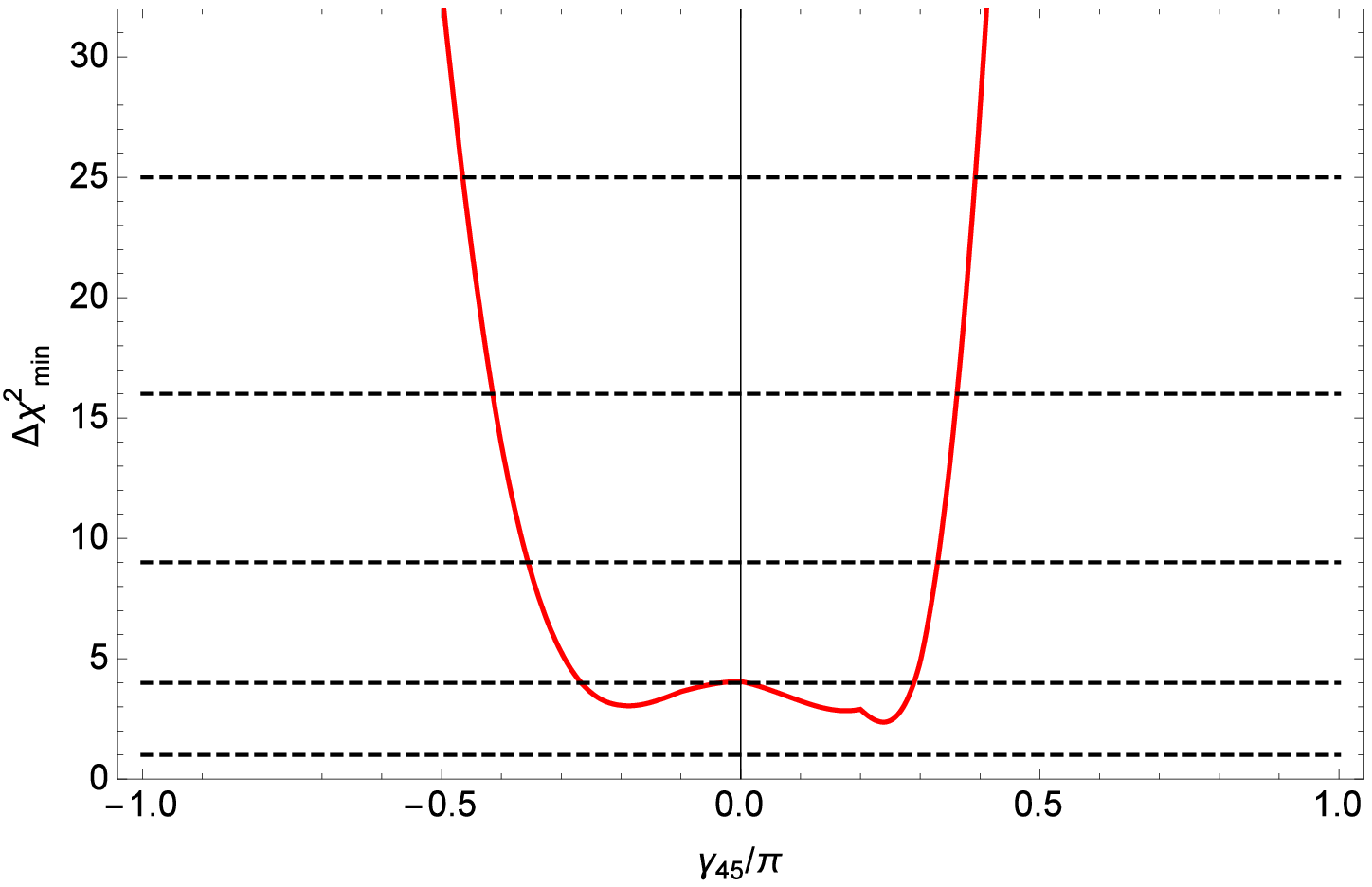}\label{fig:minChi2G45_twice_initial}} \\
	\subfigure[]{\includegraphics[width=0.45\textwidth]{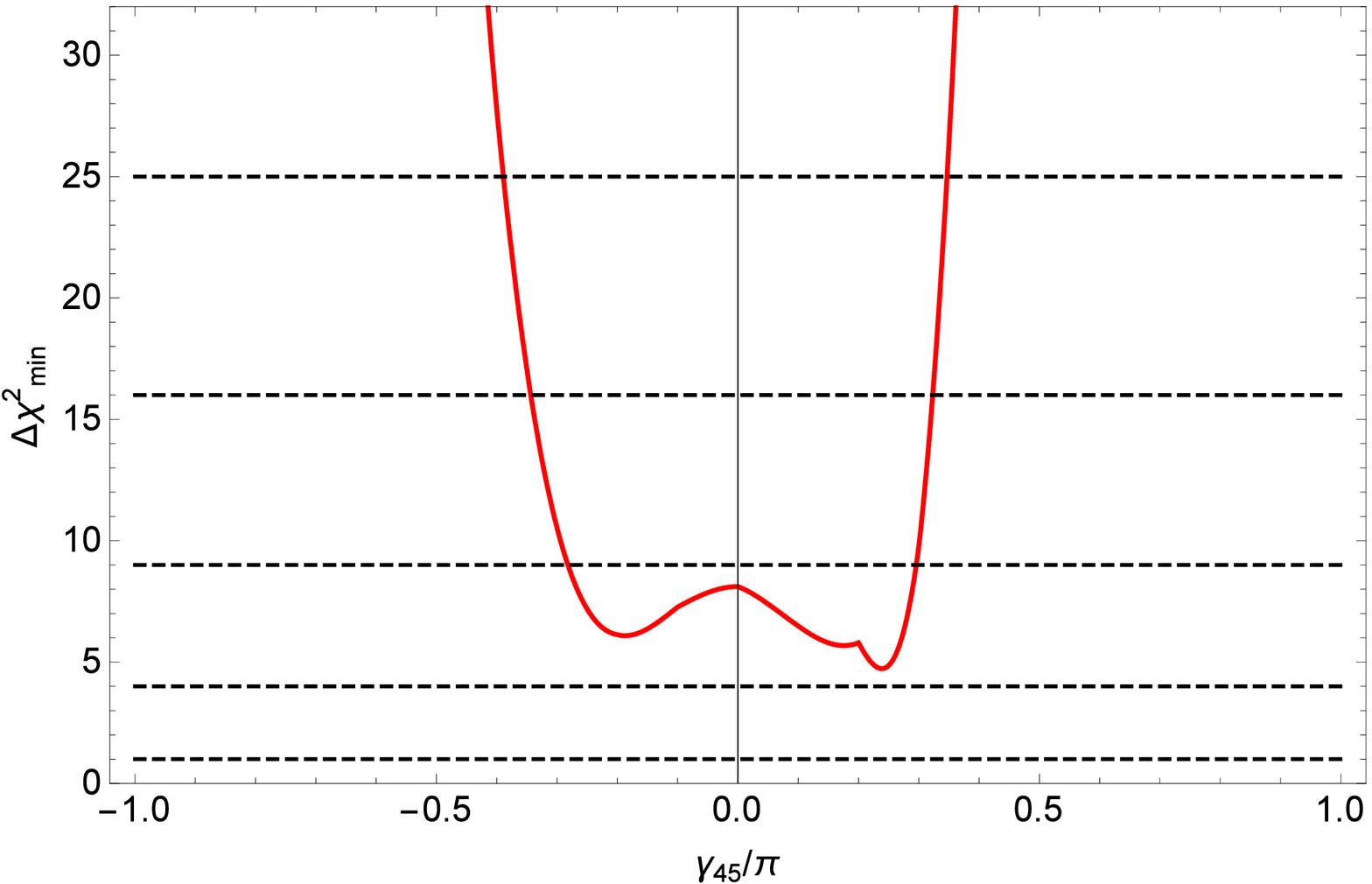}\label{fig:minChi2G45_full}}
	\subfigure[]{\includegraphics[width=0.45\textwidth]{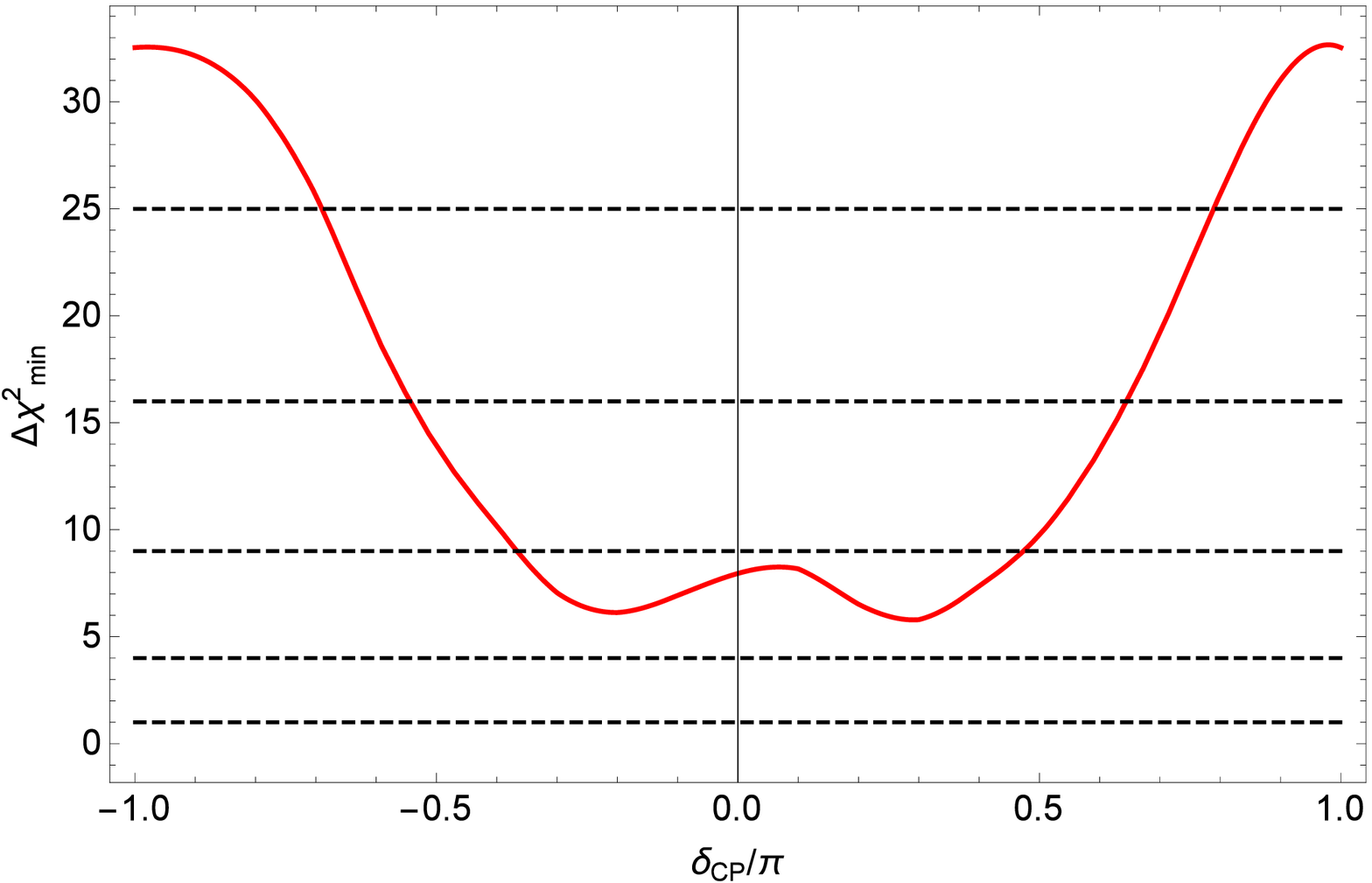}\label{fig:minChi2}}
	\caption{(Color online) (Top) Minimum $\Delta \chi^2$ for sterile neutrino discovery in LBNE for $\gamma_{45}$ at 
	10 kt (left) and 20 kt (right) after $5\nu + 5\bar{\nu}$ years of data. (Bottom) Minimum $\Delta \chi^2$ for 
	sterile neutrino discovery in 40 kt LBNE for $\gamma_{45}$ (left) and $\delta_{CP}$ (right) after 5 + 5 years 
	of data.}
\end{figure}

Large differences in the measured event rates at LBNE will result in a large $\Delta \chi^2$ value, while scenarios where there is a degeneracy between the two signals will yield a low value.  Since it is not yet known what the characteristics of  LBNE will be, we are interested in exploring at what stage the two signals could be resolved.  For this purpose we take the full design to be 40 kt with a 1.2 MW proton beam, and we estimate different initial constructions to be 10 and 20 kt with the same proton beam power. Examples of the sensitivities at the different construction stages can be seen in Figs. \ref{fig:sensitivity_initial}, \ref{fig:sensitivity_twice_initial}, and \ref{fig:sensitivity_full}.  With the initial 10 kt LBNE design we cannot fully separate the signals predicted from three-flavors and 3+2 flavors at $3\sigma$, however we can achieve that level of sensitivity with only a 20 kt detector after $5\nu + 5\bar{\nu}$ years of data taking.   

The projected $\Delta \chi^2$ values can be seen in Figs. \ref{fig:minChi2G45_initial}, \ref{fig:minChi2G45_twice_initial}, \ref{fig:minChi2G45_full}, and \ref{fig:minChi2}.  The $\Delta \chi^2$ projections are produced by finding the minimum $\Delta \chi^2$ for each $\gamma_{45}$ while allowing $\delta_{CP}$ to vary, or vice versa for the projection on $\delta_{CP}$.  It can be seen that the two signals can be distinguished at the 2 sigma level for $\gamma_{45} \lesssim -0.3\pi$ and $\gamma_{45} \gtrsim 0.3\pi$ for even the 20 kt detector size, while we can begin to distinguish the two signals at a level of $\geq 2\sigma$ for $-0.3\pi\lesssim \gamma_{45} \lesssim 0.3\pi$ with the full LBNE design after 5 + 5 years.  Likewise, the signals can only begin to be distinguished at $2\sigma$ for $-0.4 \pi \lesssim \delta_{CP} \lesssim 0.5\pi$ after 5 + 5 years with the 40 kt detector size.  Therefore if LBNE measures a CP violating phase in the normal hierarchy in this region it cannot be concluded at $\geq 3\sigma$ from LBNE data alone whether you have only three-flavors or 3+2 flavors.  In the next section we will use the results of the Daya Bay reactor experiment to place further constraints on $\gamma_{45}$ to further constrain the parameter space and help to lift this degeneracy at LBNE.

\subsection{Sensitivity in Daya Bay and T2K}
\label{sec:DB_sensitivity}

Daya Bay is a reactor experiment that measures $\bar{\nu}_e$ disappearance.  The experiment has three near detectors that measure the flux from six reactors with four different baselines, and then three far detectors that measure the final flux; recently the experiment has made a remarkable measurement of the mixing angle $\theta_{13}$ \cite{DayaBay1,DayaBay2}.  There is no sensitivity to $\delta_{CP}$, therefore we can use the Daya Bay results to constrain $\gamma_{45}$ along with $\theta_{13}$.

Complementarity between Daya Bay and LBNE can be argued from the form of the reactor oscillation probability as compared to the quantities relevant for flavor transitions found in section \ref{sec:LBNE_sensitivity}.  The Daya Bay reactor probability is given by
\begin{equation}
	\begin{aligned}
		\allowbreak
		P_{ee}\simeq 1 - &4\left\{ \left|U_{e1}\right|^2\left|U_{e3}\right|^2 \sin^2\frac{\Delta m_{31}^2 L}{4E} +
		\left|U_{e2}\right|^2\left|U_{e3}\right|^2 \sin^2\frac{\Delta m_{32}^2 L}{4E} + \left|U_{e1}\right|^2
		\left|U_{e4}\right|^2 \sin^2\frac{\Delta m_{41}^2 L}{4E} \right. \\
		& \left. + \left|U_{e2}\right|^2\left|U_{e4}\right|^2 \sin^2\frac{\Delta m_{42}^2 L}{4E} +
		\left|U_{e3}\right|^2\left|U_{e4}\right|^2 \sin^2\frac{\Delta m_{43}^2 L}{4E} + 
		\left|U_{e1}\right|^2\left|U_{e5}\right|^2 \sin^2\frac{\Delta m_{51}^2 L}{4E} \right. \\
		& \left. + \left|U_{e2}\right|^2\left|U_{e5}\right|^2 \sin^2\frac{\Delta m_{52}^2 L}{4E} +
		\left|U_{e3}\right|^2\left|U_{e5}\right|^2 \sin^2\frac{\Delta m_{53}^2 L}{4E} + 
		\left|U_{e4}\right|^2\left|U_{e5}\right|^2 \sin^2\frac{\Delta m_{54}^2 L}{4E} \right\}.
	\end{aligned}
	\label{eq:vac_osc}
\end{equation}
The oscillation terms which correspond to the large mass splittings take on their average values because the ratio $L/E$ is sufficiently large for Daya Bay.  Here we have already neglected the $\Delta m_{\textrm{sol}}^2$ term whose contribution is very small.  Taking the average values $\sin^2\frac{\Delta m_{ij}^2 L}{4E}\to \frac{1}{2}$ for $i$ or $j=4,5$, and using $\sum_i \left|U_{ei}\right|^2=1$, Eqn. \ref{eq:vac_osc} becomes
\begin{equation}
	\begin{aligned}
		P_{ee}\simeq 1 - &2\left( \left|U_{e4}\right|^2 + \left|U_{e5}\right|^2 - \left|U_{e4}\right|^4 - 
		\left|U_{e5}\right|^4 - \left|U_{e4}\right|^2 \left|U_{e5}\right|^2 \right) \\
		- &4 \left|U_{e3}\right|^2 \left( \left|U_{e1}\right|^2 + \left|U_{e2}\right|^2 \right)
		\sin^2\frac{\Delta m_{31}^2 L}{4E}\, .
	\end{aligned}
	\label{eq:vac_osc_simple}
\end{equation}
Therefore Daya Bay is sensitive to the matrix elements $U_{e1},U_{e2},U_{e3},U_{e4},U_{e5}$ and some of the mixing parameters encoded within them.  Therefore Daya Bay can help constrain elements of $U_{e2},U_{e3}$ to which LBNE is also sensitive, but is otherwise not sensitive to the same quantities as LBNE.  The intention is to allow Daya Bay to provide some additional constraints on $\gamma_{45}$ to help exclude some regions of the LBNE parameter space from section \ref{sec:LBNE_sensitivity}.

The imaginary component, $\gamma_{45}$, of the new complex sterile mixing angle has a significant control over the shaping of the probabilities for long baseline.  It was shown in \cite{Mod_Casas_Ibarra} that for long baselines when $\Delta m_{21}^2$ was negligible then the mixing elements became
\begin{equation}
	\begin{aligned}
		U_{e4} &\approx i \sqrt{\frac{m_3 m_5}{X}}e^{i(\alpha-\delta)}s_{13}\sin\left(\theta_{45}-i\gamma_{45}\right) \\
		U_{e5} &\approx i \sqrt{\frac{m_3 m_4}{X}}e^{i(\alpha-\delta)}s_{13}\cos\left(\theta_{45}-i\gamma_{45}\right) \\
	\end{aligned}
	\label{eq:mixing_elements}
\end{equation}
where $X=m_4 m_5 + m_3\frac{m_4-m_5}{2}\cos 2\theta_{45} + m_3 \frac{m_4 + m_5}{2}\cosh 2\gamma_{45}$.  The leading contributions to the Daya Bay probabilities from sterile neutrinos in Eq. \ref{eq:vac_osc_simple} lead to
\begin{equation}
	\begin{aligned}
		|U_{e4}|^2 + |U_{e5}|^2 \approx& \frac{m_3}{X}s_{13}^2 \cosh^2 2\gamma_{45} \left(m_4 \cos^2 2\theta_{45} +
		m_5 \sin^2 2\theta_{45} \right) \\
		&+ \frac{m_3}{X}s_{13}^2 \sinh^2 2\gamma_{45} \left(m_4 \sin^2 2\theta_{45} + m_5 \cos^2 2\theta_{45} \right).
	\end{aligned}
	\label{eq:DB_leading}
\end{equation}
The terms $|U_{e4}|^4,|U_{e5}|^4$ and $|U_{e4}|^2 |U_{e5}|^2$ are all $\mathcal{O}(s_{13}^4)$ and can be treated as only small corrections.  Since $m_4 \sim m_5$ then $m_4 \cos^2 2\theta_{45} + m_5 \sin^2 2\theta_{45} \approx m_4 \approx m_5$ and $m_4 \sin^2 2\theta_{45} + m_5 \cos^2 2\theta_{45} \approx m_4 \approx m_5$.  To leading order $X\sim m_4 m_5$.  Therefore the leading sterile terms relevant for Daya Bay lead to
\begin{equation}
	|U_{e4}|^2 + |U_{e5}|^2 \approx \frac{m_3}{m_4}s_{13}^2 \left(\cosh^2 2\gamma_{45} + \sinh^2 2\gamma_{45}\right).
	\label{eq:DB_leading_reduced}
\end{equation}
To leading order in the modified Casas-Ibarra parameterization $|U_{e1}|,|U_{e2}|$ and $|U_{e3}|$ are their respective quantities from three-flavor mixing.  Therefore to leading order the imaginary mixing angle component, $\gamma_{45}$, provides a strong control over probability shaping.  Since LBNE is also a long baseline experiment, the same arguments can be made to demonstrate how LBNE is sensitive to $\gamma_{45}$ in addition to the pre-existing sensitivity to $\delta_{CP}$ from the three-flavor leading order terms.  However, showing explicitly how LBNE is sensitive to $\gamma_{45}$ is complicated by the inclusion of the matter terms which renders diagonalization highly non-trivial for 5 flavors.

The energy resolution for Daya Bay is roughly $8\% / \sqrt{E(\textrm{MeV})}$, we include both statistical uncertainties as well as a $0.1\%$ uncorrelated systematic uncertainty \cite{DayaBay1,DayaBay2} in the computation of $\Delta \chi^2$.  In Fig. \ref{fig:Chi2DB} we plot the difference $\left(\Delta \chi^2\right)^{5\nu} - \left(\Delta \chi^2\right)^{3\nu}$ where $\Delta \chi^2$ is calculated by fitting to $N_{\textrm{far}}/N_{\textrm{near}}$; the projection can be seen in Fig. \ref{fig:minChi2DB}.
\begin{figure}[ht!!!]
	\centering
	\subfigure[]{\includegraphics[width=0.45\textwidth]{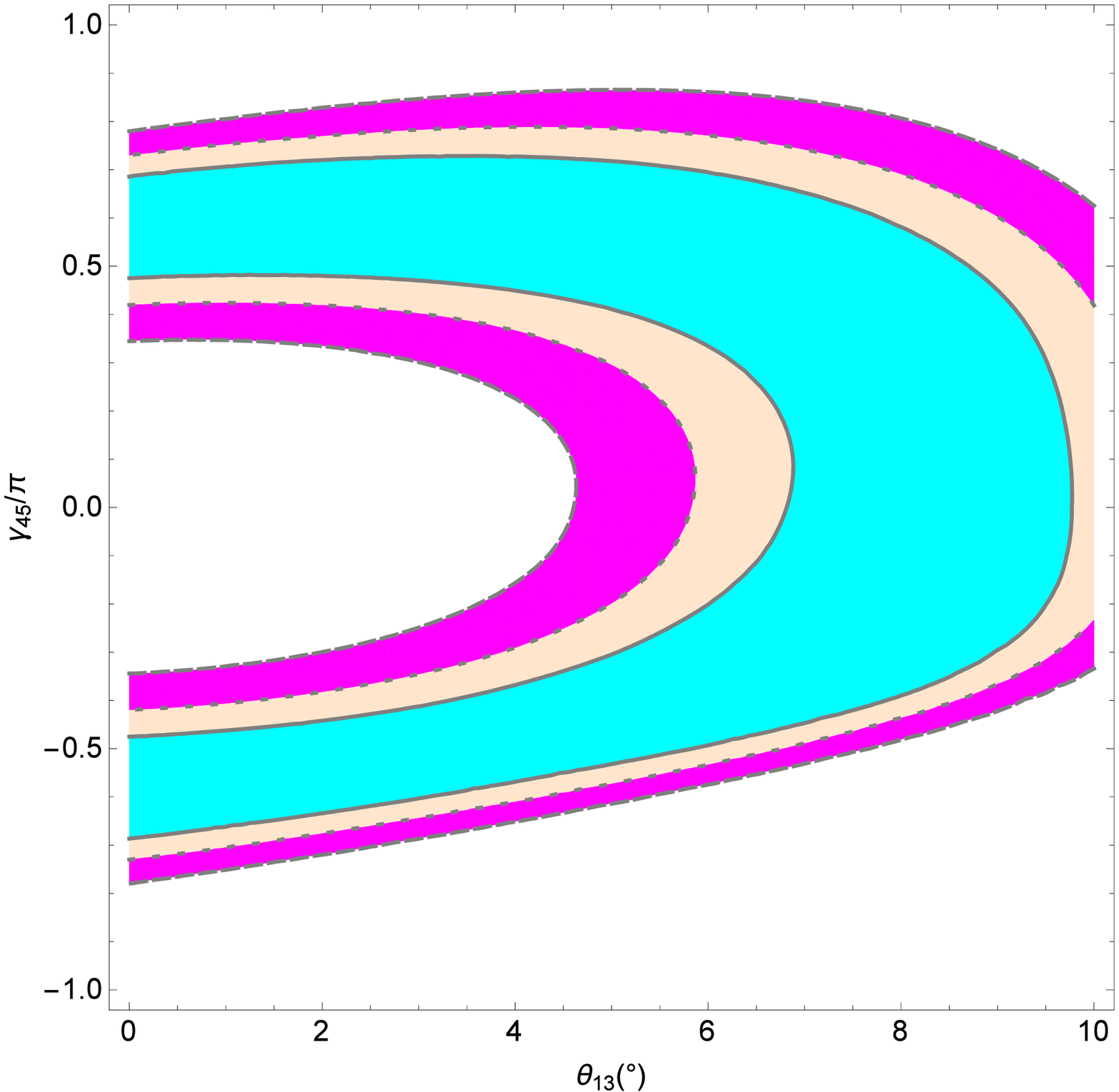}\label{fig:Chi2DB}}
	\subfigure[]{\includegraphics[width=0.45\textwidth]{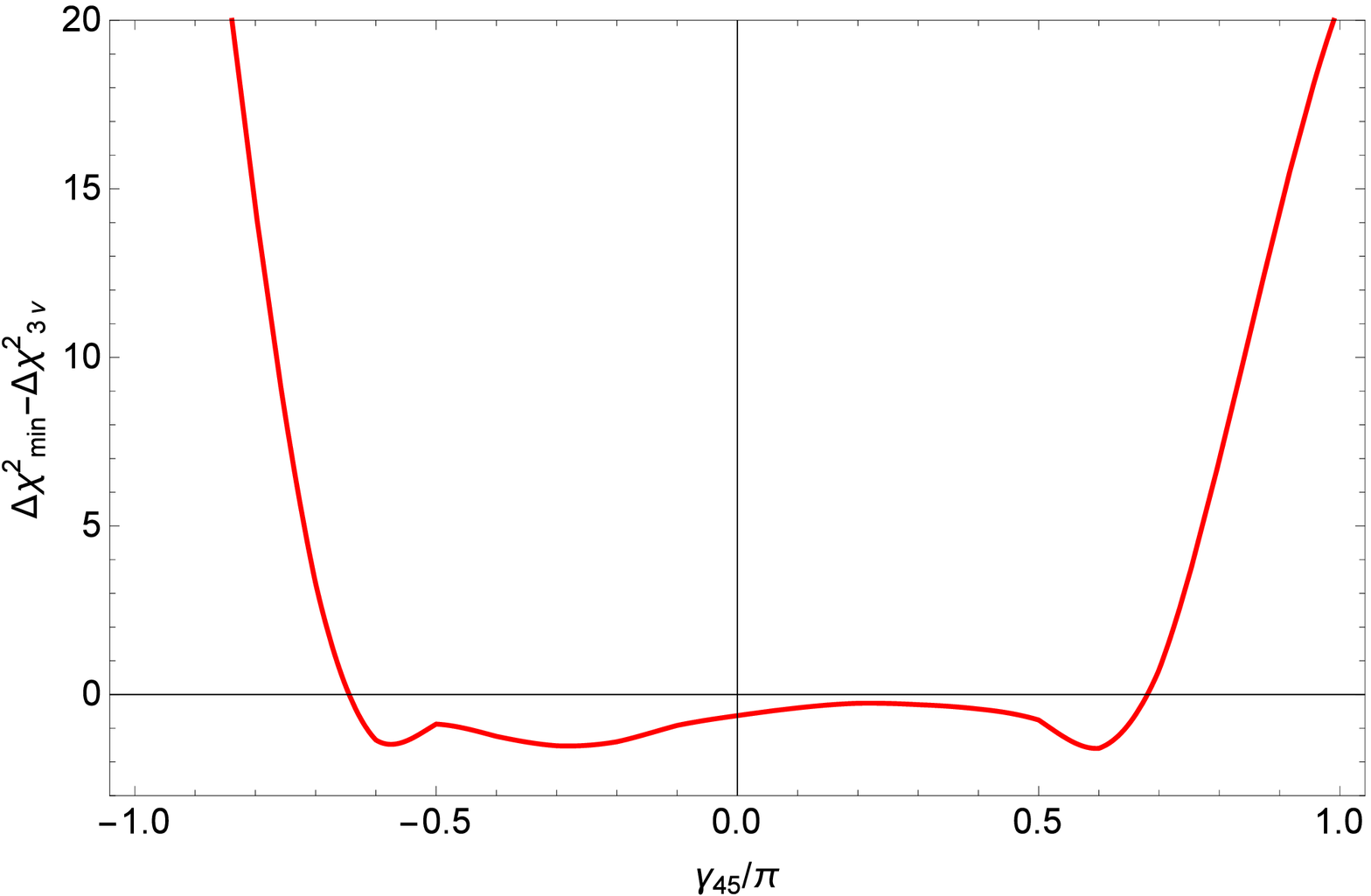}\label{fig:minChi2DB}}
	\caption{(Color online) (Left) $1\sigma (\textrm{ teal band}), 2\sigma (\textrm{ light orange band})$ and 
	$3\sigma$ (magenta band) contours of $\left(\Delta \chi^2\right)^{5\nu}-\left(\Delta \chi^2\right)^{3\nu}$, and 
	(right) the projection of the left plot.  $\Delta \chi^2$ is computed by fitting to the ratio 
	$N_{\textrm{far}}/N_{\textrm{near}}$; uncertainties included in the calculation are statistical and a 
	0.1\% uncorrelated, systematic uncertainty.  The projection is produced by minimizing $\Delta \chi^2$ over 
	the range of $\theta_{13} \in [0^{\circ}, 10^{\circ} ]$}
\end{figure}

There is a slight preference to $-0.65\pi \lesssim \gamma_{45} \lesssim 0.6\pi$, and the preference is stronger for $\gamma_{45} < 0$.  The preferred $\gamma_{45}$ values from Daya Bay helps to restrict the LBNE parameter space where the three-flavor and 3+2 flavor signals are distinguishable at $2\sigma$ in the 40 kt design, and can potentially rule out the degeneracy at $\gamma_{45} \sim 0.25\pi$.  

Recently T2K in combination with measurements of $\theta_{13}$ from Daya Bay provided results indicating sensitivity to $\delta_{CP}$ \cite{T2K}. The data from T2K can help lift the degeneracy.  T2K prefers values $\delta_{\textrm{CP}}<0$ , and can exclude the region $0.2 \lesssim \delta_{\textrm{CP}}/\pi \lesssim 0.8$ at the $90\%$ confidence level.  It is in this region excluded by T2K that we found a stronger degeneracy between the signal from the minimal sterile neutrino model and Standard Model neutrinos.  If T2K can exclude the region where degeneracies are present then it may be possible to restrict LBNE to the regions of $\delta_{CP}$ where the two signals can be distinguished. 

\begin{figure}[ht!!!]
	\centering
	\subfigure[]{\includegraphics[width=0.45\textwidth]{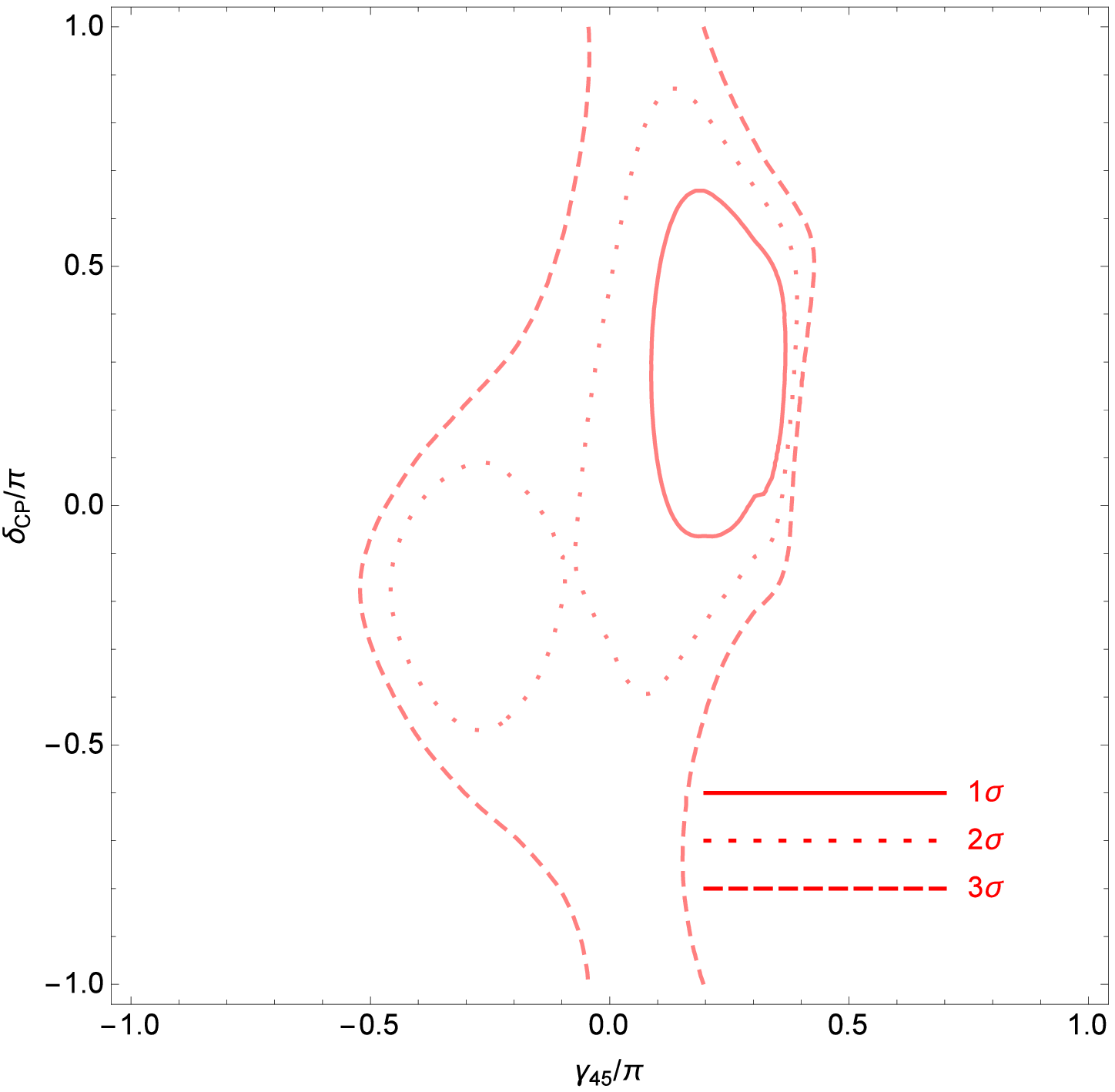}\label{fig:total_sens_quarter}}
	\subfigure[]{\includegraphics[width=0.45\textwidth]{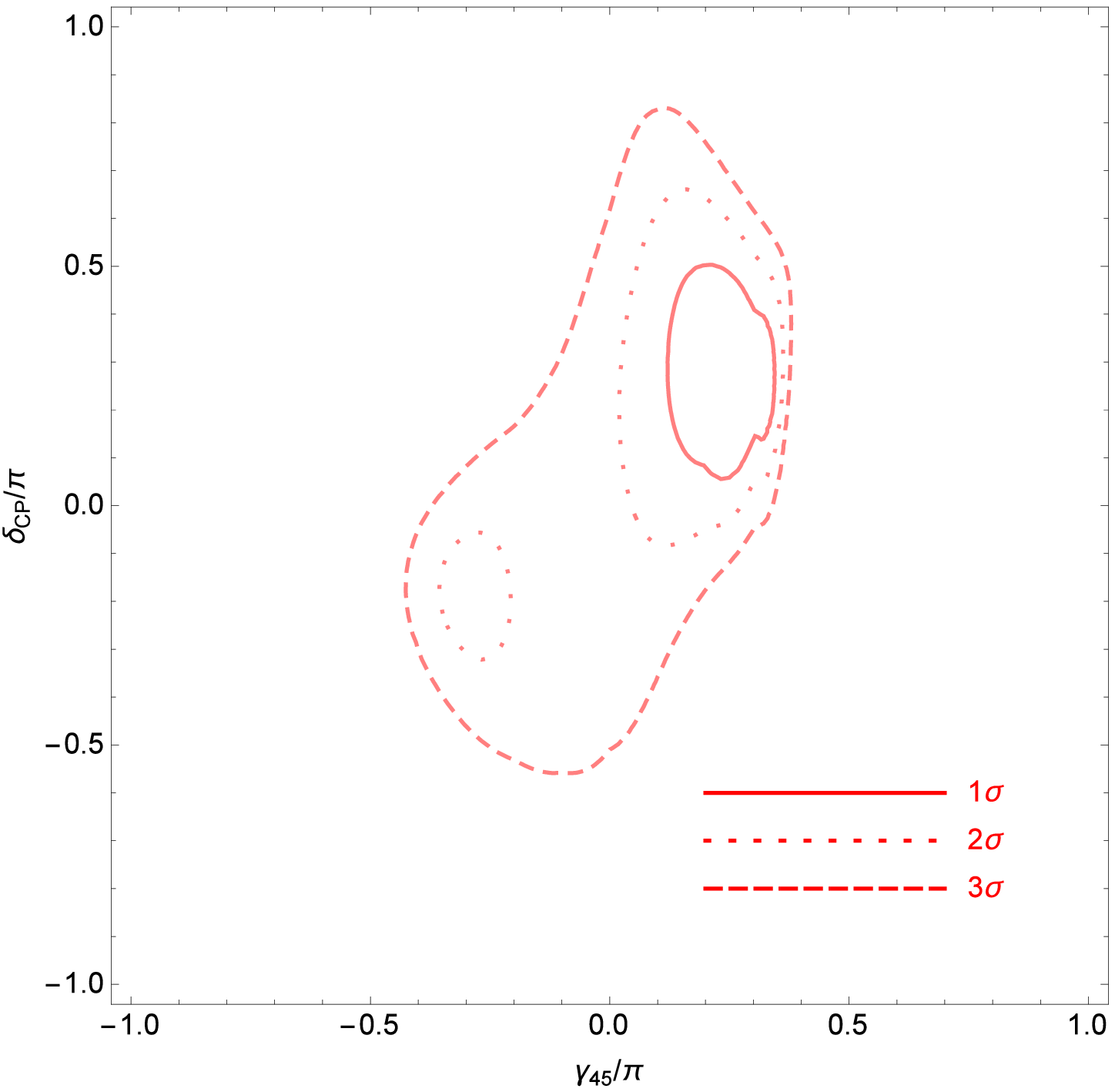}\label{fig:total_sens_half}} \\
	\subfigure[]{\includegraphics[width=0.45\textwidth]{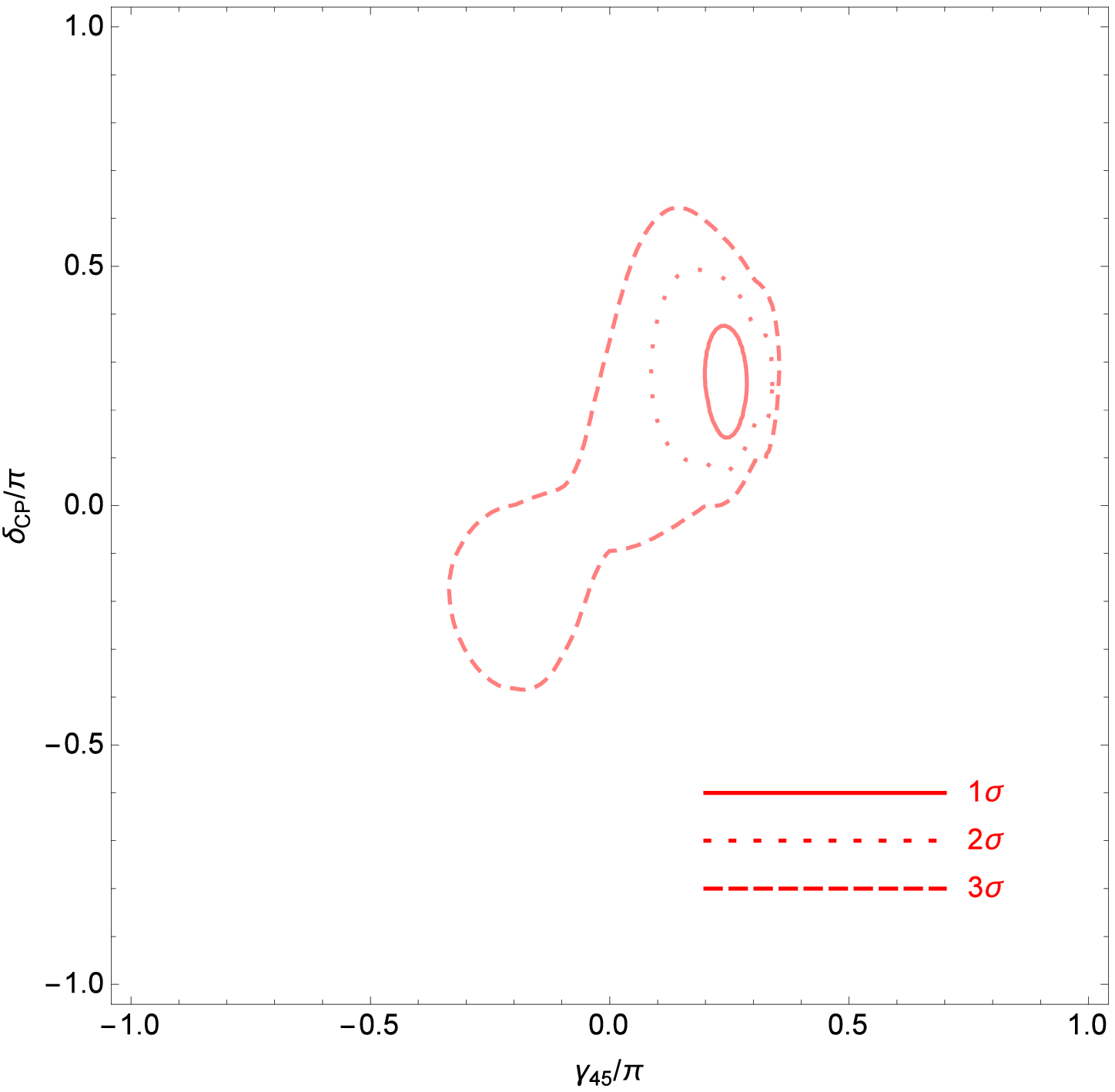}\label{fig:total_sens_full}}
	\caption{(Color online) (Top) Sensitivity contours for sterile neutrino discovery in LBNE+Daya Bay+T2K combined, with LBNE detector 
	sizes 10 kt (left), 20 kt (right), and the full, 40 kt design (bottom) all after $5\nu + 5\bar{\nu}$ years of data taking.}
\end{figure}
The full sterile neutrino sensitivity for the combined experiments LBNE+Daya Bay+T2K can be seen in Figs. \ref{fig:total_sens_quarter}, \ref{fig:total_sens_half} and \ref{fig:total_sens_full}.  Comparing the combined sensitivities to those for LBNE only in Figs. \ref{fig:sensitivity_initial}, \ref{fig:sensitivity_twice_initial}, and \ref{fig:sensitivity_full} it can be seen that the $1\sigma$ and $2\sigma$ contours are noticeably altered and in certain cases significantly contracted, however sensitivities at the $3\sigma$ level remain mostly the same as before.  For the 10 kt LBNE detector size we still cannot completely distinguish between signals predicted by 3 flavors and 3+2 flavors at a $3\sigma$ level.  However we can achieve that level at larger detector sizes and furthermore we can rule out more of the parameter space of this 3+2 minimal model as compared with LBNE only.  For example, with the full 40 kt detector sensitivities at the $2\sigma$ level vanish for $\gamma_{45} < 0$ and $\delta_{CP} < 0$, and the $3\sigma$ level is pinched near $\gamma_{45}, \delta_{CP} \sim 0$.  Although the $3\sigma$ sensitivities remain mostly unchanged, the altered $1\sigma$ and $2\sigma$ contours indicate larger asymmetries between the regions separated by $\gamma_{45} = 0$ and $\delta_{CP} = 0$ for the combined data compared to only LBNE.
\begin{figure}[ht!!!]
	\centering
	\subfigure[]{\includegraphics[width=0.45\textwidth]{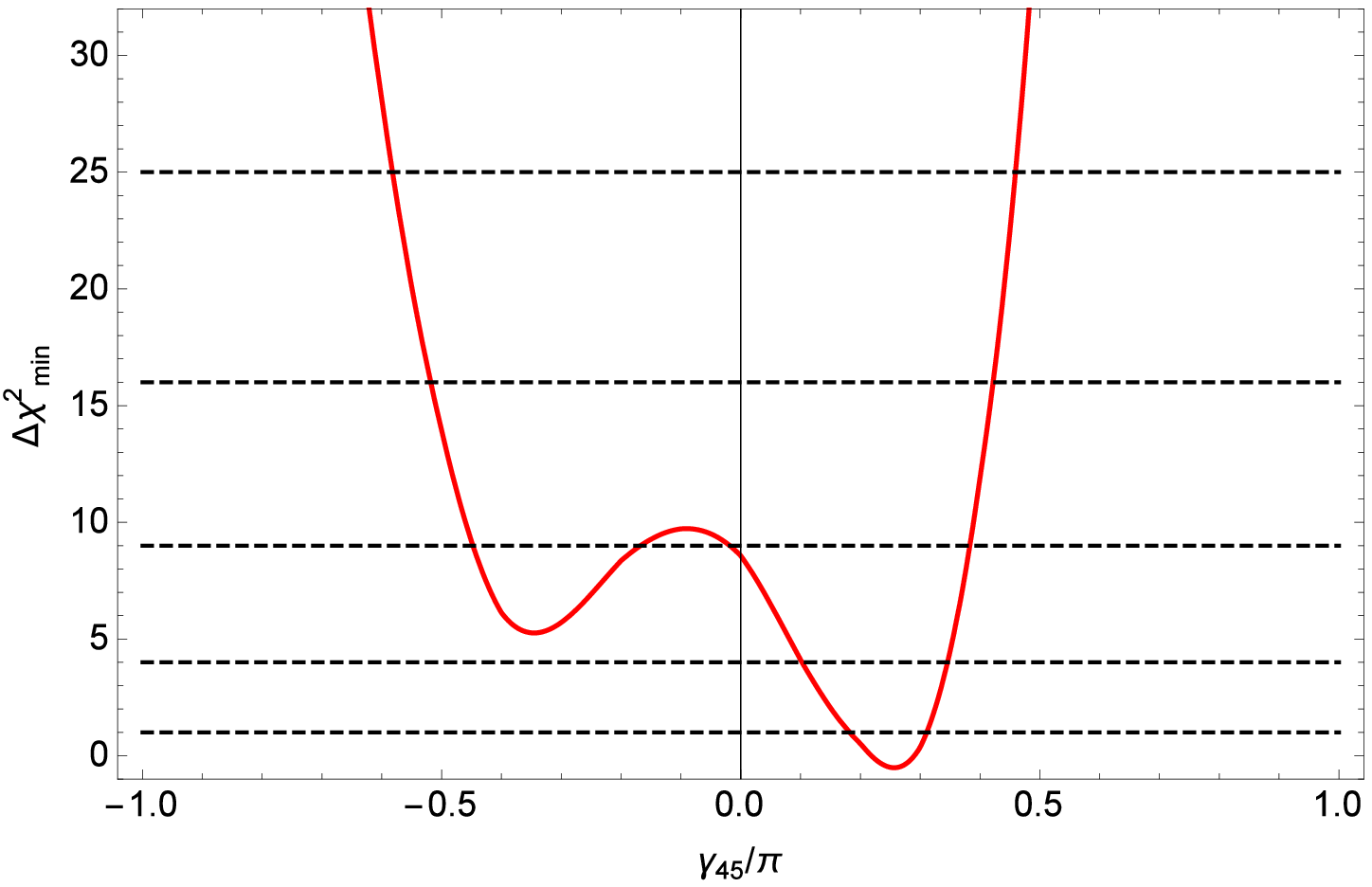}\label{fig:total_sens_quarter_min}}
	\subfigure[]{\includegraphics[width=0.45\textwidth]{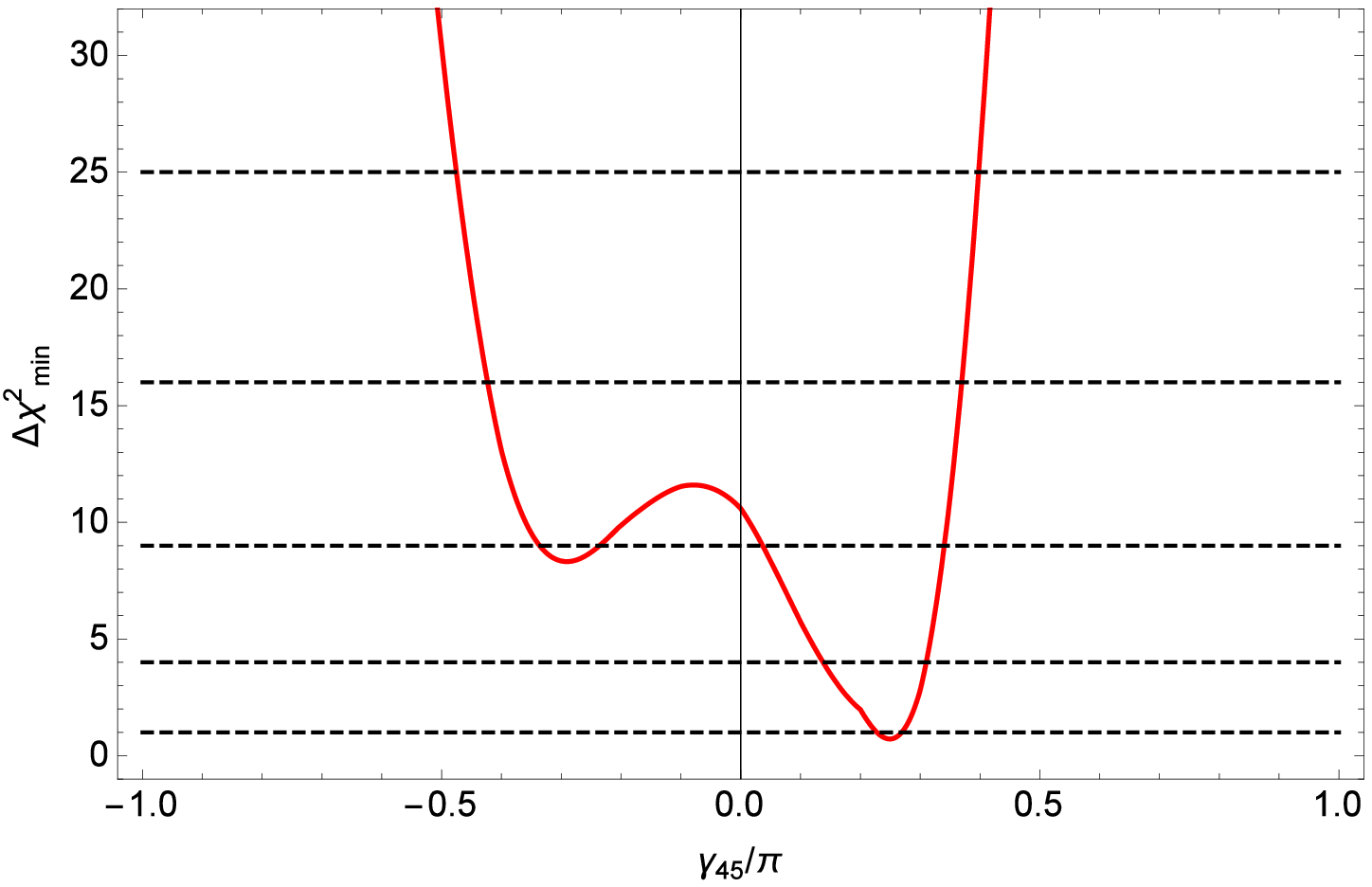}\label{fig:total_sens_half_min}} \\
	\subfigure[]{\includegraphics[width=0.45\textwidth]{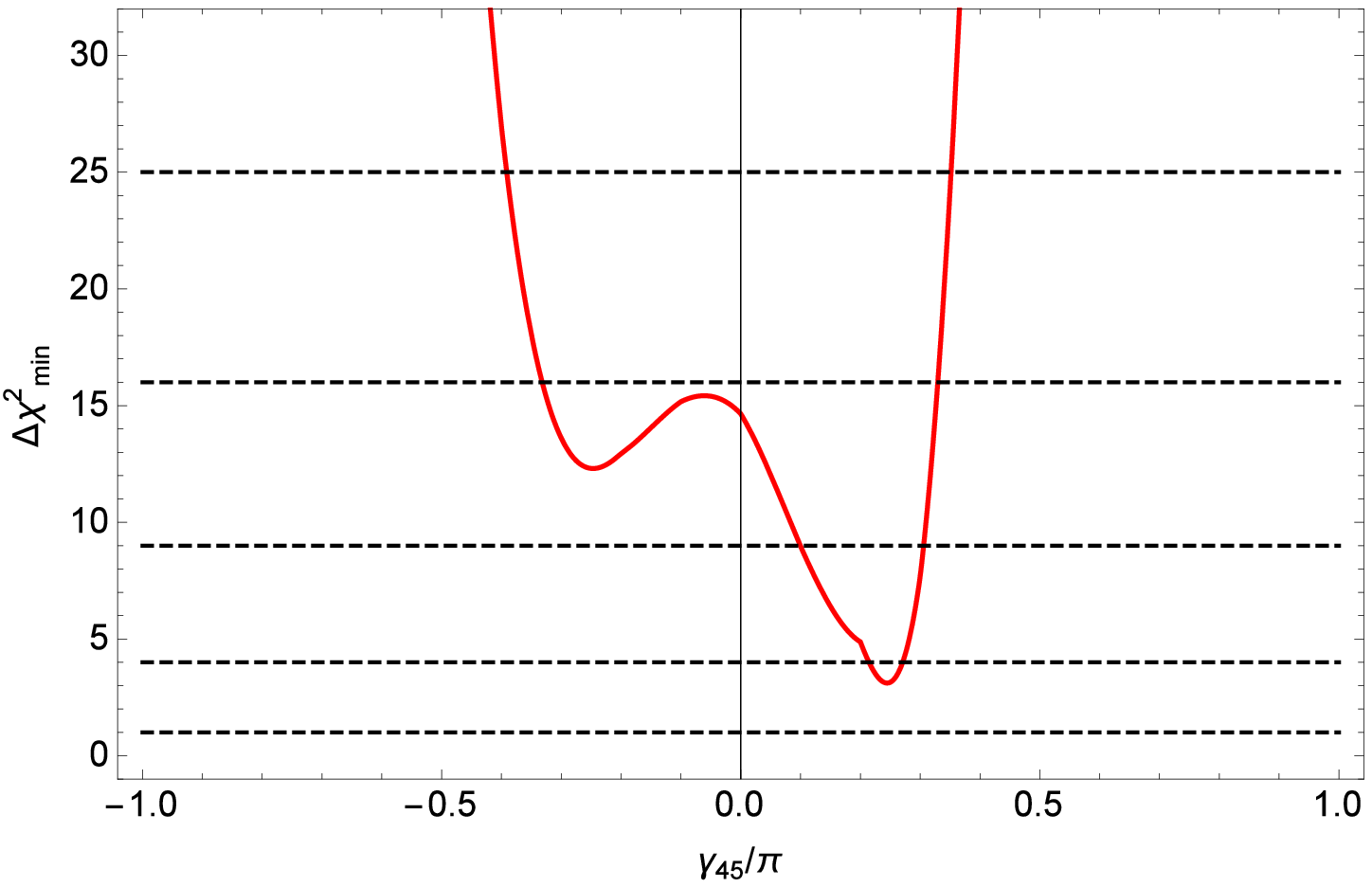}\label{fig:total_sens_full_min}} 
	\subfigure[]{\includegraphics[width=0.45\textwidth]{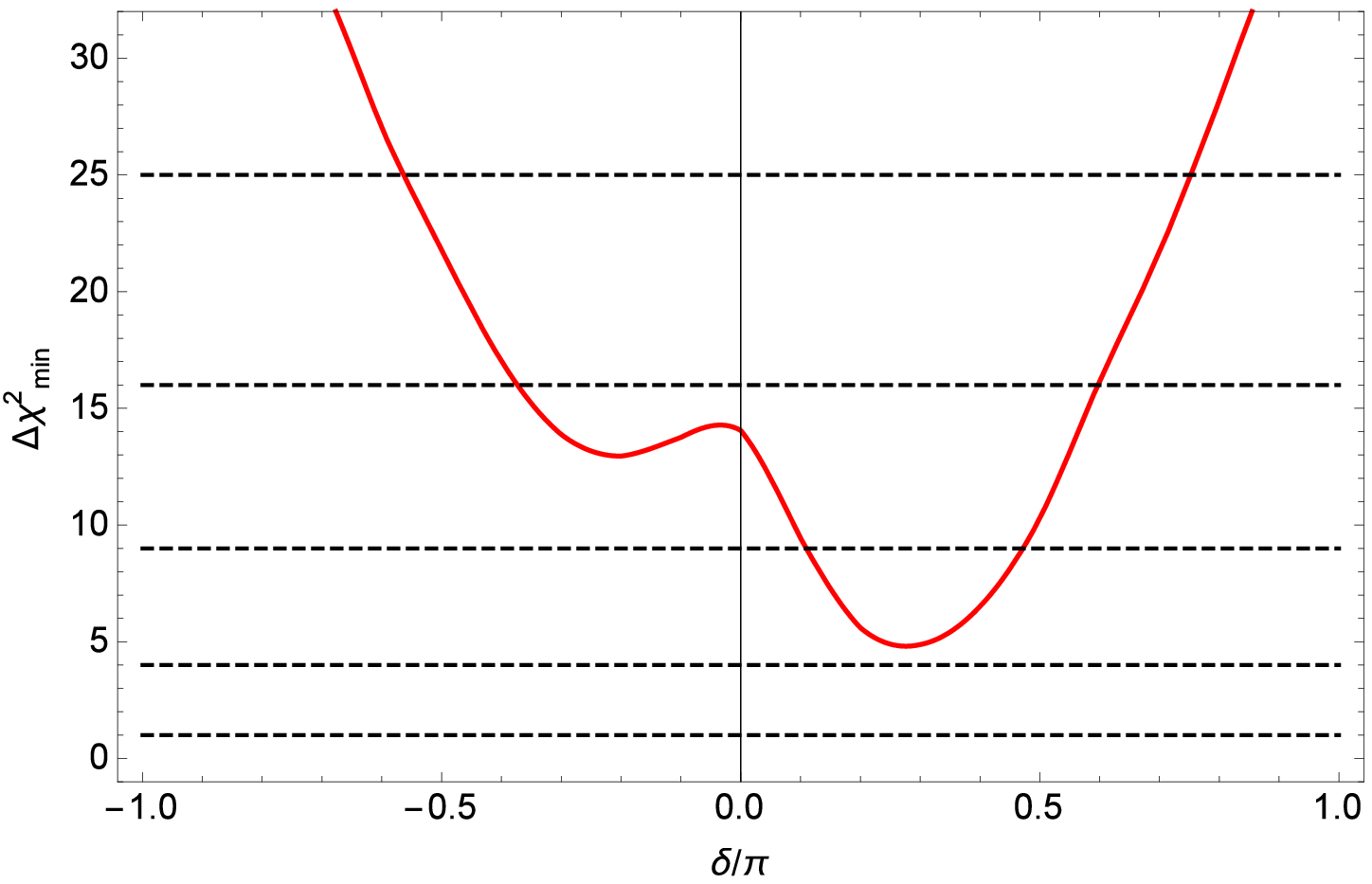}\label{fig:total_sens_full_min_delta}} 
	\caption{(Color online) (Top) Minimum $\Delta \chi^2$ for sterile neutrino discovery in LBNE+Daya Bay+T2K combined, with LBNE 
	detector sizes 10 kt (top), 20 kt (middle).  (Bottom) Minimum $\Delta \chi^2$ for sterile neutrino discovery in LBNE+Daya Bay+T2K combined, 
	with the full LBNE detector size, 40 kt design for $\gamma_{45}$ (left) and $\delta_{CP}$ (right) all after $5\nu + 5\bar{\nu}$ years 
	of data taking.}
\end{figure}

The projected $\Delta \chi^2$ can be seen in Figs. \ref{fig:total_sens_quarter_min}, \ref{fig:total_sens_half_min}, \ref{fig:total_sens_full_min} and \ref{fig:total_sens_full_min_delta}.  It can be seen that even with the 10 kt LBNE detector size combining the results from LBNE, Daya Bay and T2K we are able to distinguish the signals predicted by 3 flavors and 3+2 flavors at a level $> 2\sigma$ for all negative $\gamma_{45}$ parameter values.  With the 40 kt LBNE design we can make the distinction at a level $>3\sigma$ for all $\gamma_{45}<0$ and $\delta_{CP}<0$, and at a level $>2\sigma$ for $\delta_{CP}>0$.  However for $0.1\pi \leq \gamma_{45} \lesssim 0.35\pi$ the two signals cannot be separated at the $2\sigma$ level in the 10 kt detector, and even with the 40 kt design the signals cannot be separated at the $2\sigma$ level for parameter values near $\gamma_{45} =0.25\pi$.

While the LBNE near detector may add very useful information, we have not considered its effects in this analysis.  A full design of a near detector does not yet exist and any results would strongly depend on the detector setup and characteristics; in addition, such an analysis would need a good understanding of the beam, which is likely not precise enough.  Furthermore, a near detector analysis will be limited by systematic uncertainties and would likely be too speculative at this time, though studies involving information provided by a LBNE near detector may be needed in the future.

	\section{Concluding remarks}
\label{sec:Conclusion}

In addition to the stated goal of measuring the CP-violating phase for three-flavor neutrino oscillations, LBNE is also poised to potentially measure signals from new physics involving sterile neutrinos.  We examined a particular parameterization for a model involving two new sterile neutrino flavors where LBNE is sensitive to the imaginary component of the new mixing angle ($\gamma_{45}$).  Having examined the $\nu_{\mu} \to \nu_e$ oscillation probabilities for different parameter values of $\delta_{CP}$ and $\gamma_{45}$ it has been shown that there are regions where LBNE can clearly resolve the new physics and Standard Model physics signals.  Furthermore it has been shown that with the 40 kt detector design in place this measurement is possible above $2\sigma$ over the entire 3+2 parameter space after 5 + 5 years of data taking.  Once LBNE begins to take data it can provide further evidence for the existence of sterile neutrinos or help rule out this particular model.

Best fits from \cite{Mod_Casas_Ibarra} indicate that $\gamma_{45} \approx -1.15$ which is in the region where the two signals are clearly different.  There are degeneracies between the signals for a small region in parameter space, however by combing LBNE data with restrictions on $\gamma_{45}$ from Daya Bay, as well as the current $\delta_{CP}$ exclusions from T2K, it is possible to lift this degeneracy and evince or help rule out this particular model of sterile neutrinos.
	
	\bigskip
	\noindent
	{\Large \bf Acknowledgments}
	
	\noindent 
	This work was supported in part by the US Department of Energy under contract DE-SC0010534 and by the National Science Foundation under Grant No. NSF PHY11-25915.

\endgroup

\end{document}